\documentclass[letterpaper]{article} % DO NOT CHANGE THIS
\usepackage{aaai24}  % DO NOT CHANGE THIS
\usepackage{times}  % DO NOT CHANGE THIS
\usepackage{helvet}  % DO NOT CHANGE THIS
\usepackage{courier}  % DO NOT CHANGE THIS
\usepackage[hyphens]{url}  % DO NOT CHANGE THIS
\usepackage{graphicx} % DO NOT CHANGE THIS
\usepackage{natbib}  % DO NOT CHANGE THIS AND DO NOT ADD ANY OPTIONS TO IT
\usepackage{caption} % DO NOT CHANGE THIS AND DO NOT ADD ANY OPTIONS TO IT
\usepackage{algorithm}
\usepackage{algorithmic}

\usepackage{newfloat}
\usepackage{listings}
\DeclareCaptionStyle{ruled}{labelfont=normalfont,labelsep=colon,strut=off} % DO NOT CHANGE THIS
\floatstyle{ruled}
\newfloat{listing}{tb}{lst}{}
\floatname{listing}{Listing}
\title{Disjoint Partial Enumeration without Blocking Clauses}
\author{
    Giuseppe Spallitta\textsuperscript{\rm 1}, 
    Roberto Sebastiani\textsuperscript{\rm 1}, 
    Armin Biere\textsuperscript{\rm 2}
}
\affiliations{
    \textsuperscript{\rm 1}DISI, University of Trento\\
    \textsuperscript{\rm 2}University of Freiburg\\
    giuseppe.spallitta@unitn.it, roberto.sebastiani@unitn.it, biere@cs.uni-freiburg.de
}

\usepackage{xcolor}
\usepackage{xspace}
\usepackage{amsfonts}
\usepackage{amsmath}
\usepackage{caption}
\usepackage{subcaption}

\usepackage{bibentry}
\definecolor {snow}                {rgb}{1.00,0.98,0.98}
\definecolor {ghostwhite}          {rgb}{0.97,0.97,1.00}
\definecolor {whitesmoke}          {rgb}{0.96,0.96,0.96}
\definecolor {gainsboro}           {rgb}{0.86,0.86,0.86}
\definecolor {floralwhite}         {rgb}{1.00,0.98,0.94}
\definecolor {oldlace}             {rgb}{0.99,0.96,0.90}
\definecolor {linen}               {rgb}{0.98,0.94,0.90}
\definecolor {antiquewhite}        {rgb}{0.98,0.92,0.84}
\definecolor {papayawhip}          {rgb}{1.00,0.94,0.84}
\definecolor {blanchedalmond}      {rgb}{1.00,0.92,0.80}
\definecolor {bisque}              {rgb}{1.00,0.89,0.77}
\definecolor {peachpuff}           {rgb}{1.00,0.85,0.73}
\definecolor {navajowhite}         {rgb}{1.00,0.87,0.68}
\definecolor {moccasin}            {rgb}{1.00,0.89,0.71}
\definecolor {cornsilk}            {rgb}{1.00,0.97,0.86}
\definecolor {ivory}               {rgb}{1.00,1.00,0.94}
\definecolor {lemonchiffon}        {rgb}{1.00,0.98,0.80}
\definecolor {seashell}            {rgb}{1.00,0.96,0.93}
\definecolor {honeydew}            {rgb}{0.94,1.00,0.94}
\definecolor {mintcream}           {rgb}{0.96,1.00,0.98}
\definecolor {azure}               {rgb}{0.94,1.00,1.00}
\definecolor {aliceblue}           {rgb}{0.94,0.97,1.00}
\definecolor {lavender}            {rgb}{0.90,0.90,0.98}
\definecolor {lavenderblush}       {rgb}{1.00,0.94,0.96}
\definecolor {mistyrose}           {rgb}{1.00,0.89,0.88}
\definecolor {white}               {rgb}{1.00,1.00,1.00}
\definecolor {black}               {rgb}{0.00,0.00,0.00}
\definecolor {darkslategray}       {rgb}{0.18,0.31,0.31}
\definecolor {dimgray}             {rgb}{0.41,0.41,0.41}
\definecolor {slategray}           {rgb}{0.44,0.50,0.56}
\definecolor {lightslategray}      {rgb}{0.47,0.53,0.60}
\definecolor {gray}                {rgb}{0.75,0.75,0.75}
\definecolor {lightgrey}           {rgb}{0.83,0.83,0.83}
\definecolor {midnightblue}        {rgb}{0.10,0.10,0.44}
\definecolor {navy}                {rgb}{0.00,0.00,0.50}
\definecolor {cornflowerblue}      {rgb}{0.39,0.58,0.93}
\definecolor {darkslateblue}       {rgb}{0.28,0.24,0.55}
\definecolor {slateblue}           {rgb}{0.42,0.35,0.80}
\definecolor {mediumslateblue}     {rgb}{0.48,0.41,0.93}
\definecolor {lightslateblue}      {rgb}{0.52,0.44,1.00}
\definecolor {mediumblue}          {rgb}{0.00,0.00,0.80}
\definecolor {royalblue}           {rgb}{0.25,0.41,0.88}
\definecolor {blue}                {rgb}{0.00,0.00,1.00}
\definecolor {dodgerblue}          {rgb}{0.12,0.56,1.00}
\definecolor {deepskyblue}         {rgb}{0.00,0.75,1.00}
\definecolor {skyblue}             {rgb}{0.53,0.81,0.92}
\definecolor {lightskyblue}        {rgb}{0.53,0.81,0.98}
\definecolor {steelblue}           {rgb}{0.27,0.51,0.71}
\definecolor {lightsteelblue}      {rgb}{0.69,0.77,0.87}
\definecolor {lightblue}           {rgb}{0.68,0.85,0.90}
\definecolor {powderblue}          {rgb}{0.69,0.88,0.90}
\definecolor {paleturquoise}       {rgb}{0.69,0.93,0.93}
\definecolor {darkturquoise}       {rgb}{0.00,0.81,0.82}
\definecolor {mediumturquoise}     {rgb}{0.28,0.82,0.80}
\definecolor {turquoise}           {rgb}{0.25,0.88,0.82}
\definecolor {cyan}                {rgb}{0.00,1.00,1.00}
\definecolor {lightcyan}           {rgb}{0.88,1.00,1.00}
\definecolor {cadetblue}           {rgb}{0.37,0.62,0.63}
\definecolor {mediumaquamarine}    {rgb}{0.40,0.80,0.67}
\definecolor {aquamarine}          {rgb}{0.50,1.00,0.83}
\definecolor {darkgreen}           {rgb}{0.00,0.39,0.00}
\definecolor {darkolivegreen}      {rgb}{0.33,0.42,0.18}
\definecolor {darkseagreen}        {rgb}{0.56,0.74,0.56}
\definecolor {seagreen}            {rgb}{0.18,0.55,0.34}
\definecolor {mediumseagreen}      {rgb}{0.24,0.70,0.44}
\definecolor {lightseagreen}       {rgb}{0.13,0.70,0.67}
\definecolor {palegreen}           {rgb}{0.60,0.98,0.60}
\definecolor {springgreen}         {rgb}{0.00,1.00,0.50}
\definecolor {lawngreen}           {rgb}{0.49,0.99,0.00}
\definecolor {green}               {rgb}{0.00,1.00,0.00}
\definecolor {chartreuse}          {rgb}{0.50,1.00,0.00}
\definecolor {mediumspringgreen}   {rgb}{0.00,0.98,0.60}
\definecolor {greenyellow}         {rgb}{0.68,1.00,0.18}
\definecolor {limegreen}           {rgb}{0.20,0.80,0.20}
\definecolor {yellowgreen}         {rgb}{0.60,0.80,0.20}
\definecolor {forestgreen}         {rgb}{0.13,0.55,0.13}
\definecolor {olivedrab}           {rgb}{0.42,0.56,0.14}
\definecolor {darkkhaki}           {rgb}{0.74,0.72,0.42}
\definecolor {khaki}               {rgb}{0.94,0.90,0.55}
\definecolor {palegoldenrod}       {rgb}{0.93,0.91,0.67}
\definecolor {lightgoldenrodyellow} {rgb}{0.98,0.98,0.82}
\definecolor {lightyellow}         {rgb}{1.00,1.00,0.88}
\definecolor {yellow}              {rgb}{1.00,1.00,0.00}
\definecolor {gold}                {rgb}{1.00,0.84,0.00}
\definecolor {lightgoldenrod}      {rgb}{0.93,0.87,0.51}
\definecolor {goldenrod}           {rgb}{0.85,0.65,0.13}
\definecolor {darkgoldenrod}       {rgb}{0.72,0.53,0.04}
\definecolor {rosybrown}           {rgb}{0.74,0.56,0.56}
\definecolor {indianred}           {rgb}{0.80,0.36,0.36}
\definecolor {saddlebrown}         {rgb}{0.55,0.27,0.07}
\definecolor {sienna}              {rgb}{0.63,0.32,0.18}
\definecolor {peru}                {rgb}{0.80,0.52,0.25}
\definecolor {burlywood}           {rgb}{0.87,0.72,0.53}
\definecolor {beige}               {rgb}{0.96,0.96,0.86}
\definecolor {wheat}               {rgb}{0.96,0.87,0.70}
\definecolor {sandybrown}          {rgb}{0.96,0.64,0.38}
\definecolor {tan}                 {rgb}{0.82,0.71,0.55}
\definecolor {chocolate}           {rgb}{0.82,0.41,0.12}
\definecolor {firebrick}           {rgb}{0.70,0.13,0.13}
\definecolor {brown}               {rgb}{0.65,0.16,0.16}
\definecolor {darksalmon}          {rgb}{0.91,0.59,0.48}
\definecolor {salmon}              {rgb}{0.98,0.50,0.45}
\definecolor {lightsalmon}         {rgb}{1.00,0.63,0.48}
\definecolor {orange}              {rgb}{1.00,0.65,0.00}
\definecolor {darkorange}          {rgb}{1.00,0.55,0.00}
\definecolor {coral}               {rgb}{1.00,0.50,0.31}
\definecolor {lightcoral}          {rgb}{0.94,0.50,0.50}
\definecolor {tomato}              {rgb}{1.00,0.39,0.28}
\definecolor {orangered}           {rgb}{1.00,0.27,0.00}
\definecolor {red}                 {rgb}{1.00,0.00,0.00}
\definecolor {hotpink}             {rgb}{1.00,0.41,0.71}
\definecolor {deeppink}            {rgb}{1.00,0.08,0.58}
\definecolor {pink}                {rgb}{1.00,0.75,0.80}
\definecolor {lightpink}           {rgb}{1.00,0.71,0.76}
\definecolor {palevioletred}       {rgb}{0.86,0.44,0.58}
\definecolor {maroon}              {rgb}{0.69,0.19,0.38}
\definecolor {mediumvioletred}     {rgb}{0.78,0.08,0.52}
\definecolor {violetred}           {rgb}{0.82,0.13,0.56}
\definecolor {magenta}             {rgb}{1.00,0.00,1.00}
\definecolor {violet}              {rgb}{0.93,0.51,0.93}
\definecolor {plum}                {rgb}{0.87,0.63,0.87}
\definecolor {orchid}              {rgb}{0.85,0.44,0.84}
\definecolor {mediumorchid}        {rgb}{0.73,0.33,0.83}
\definecolor {darkorchid}          {rgb}{0.60,0.20,0.80}
\definecolor {darkviolet}          {rgb}{0.58,0.00,0.83}
\definecolor {blueviolet}          {rgb}{0.54,0.17,0.89}
\definecolor {purple}              {rgb}{0.63,0.13,0.94}
\definecolor {mediumpurple}        {rgb}{0.58,0.44,0.86}
\definecolor {thistle}             {rgb}{0.85,0.75,0.85}
\definecolor {snow2}               {rgb}{0.93,0.91,0.91}
\definecolor {snow3}               {rgb}{0.80,0.79,0.79}
\definecolor {snow4}               {rgb}{0.55,0.54,0.54}
\definecolor {seashell2}           {rgb}{0.93,0.90,0.87}
\definecolor {seashell3}           {rgb}{0.80,0.77,0.75}
\definecolor {seashell4}           {rgb}{0.55,0.53,0.51}
\definecolor {antiquewhite1}       {rgb}{1.00,0.94,0.86}
\definecolor {antiquewhite2}       {rgb}{0.93,0.87,0.80}
\definecolor {antiquewhite3}       {rgb}{0.80,0.75,0.69}
\definecolor {antiquewhite4}       {rgb}{0.55,0.51,0.47}
\definecolor {bisque2}             {rgb}{0.93,0.84,0.72}
\definecolor {bisque3}             {rgb}{0.80,0.72,0.62}
\definecolor {bisque4}             {rgb}{0.55,0.49,0.42}
\definecolor {peachpuff2}          {rgb}{0.93,0.80,0.68}
\definecolor {peachpuff3}          {rgb}{0.80,0.69,0.58}
\definecolor {peachpuff4}          {rgb}{0.55,0.47,0.40}
\definecolor {navajowhite2}        {rgb}{0.93,0.81,0.63}
\definecolor {navajowhite3}        {rgb}{0.80,0.70,0.55}
\definecolor {navajowhite4}        {rgb}{0.55,0.47,0.37}
\definecolor {lemonchiffon2}       {rgb}{0.93,0.91,0.75}
\definecolor {lemonchiffon3}       {rgb}{0.80,0.79,0.65}
\definecolor {lemonchiffon4}       {rgb}{0.55,0.54,0.44}
\definecolor {cornsilk2}           {rgb}{0.93,0.91,0.80}
\definecolor {cornsilk3}           {rgb}{0.80,0.78,0.69}
\definecolor {cornsilk4}           {rgb}{0.55,0.53,0.47}
\definecolor {ivory2}              {rgb}{0.93,0.93,0.88}
\definecolor {ivory3}              {rgb}{0.80,0.80,0.76}
\definecolor {ivory4}              {rgb}{0.55,0.55,0.51}
\definecolor {honeydew2}           {rgb}{0.88,0.93,0.88}
\definecolor {honeydew3}           {rgb}{0.76,0.80,0.76}
\definecolor {honeydew4}           {rgb}{0.51,0.55,0.51}
\definecolor {lavenderblush2}      {rgb}{0.93,0.88,0.90}
\definecolor {lavenderblush3}      {rgb}{0.80,0.76,0.77}
\definecolor {lavenderblush4}      {rgb}{0.55,0.51,0.53}
\definecolor {mistyrose2}          {rgb}{0.93,0.84,0.82}
\definecolor {mistyrose3}          {rgb}{0.80,0.72,0.71}
\definecolor {mistyrose4}          {rgb}{0.55,0.49,0.48}
\definecolor {azure2}              {rgb}{0.88,0.93,0.93}
\definecolor {azure3}              {rgb}{0.76,0.80,0.80}
\definecolor {azure4}              {rgb}{0.51,0.55,0.55}
\definecolor {slateblue1}          {rgb}{0.51,0.44,1.00}
\definecolor {slateblue2}          {rgb}{0.48,0.40,0.93}
\definecolor {slateblue3}          {rgb}{0.41,0.35,0.80}
\definecolor {slateblue4}          {rgb}{0.28,0.24,0.55}
\definecolor {royalblue1}          {rgb}{0.28,0.46,1.00}
\definecolor {royalblue2}          {rgb}{0.26,0.43,0.93}
\definecolor {royalblue3}          {rgb}{0.23,0.37,0.80}
\definecolor {royalblue4}          {rgb}{0.15,0.25,0.55}
\definecolor {blue2}               {rgb}{0.00,0.00,0.93}
\definecolor {blue4}               {rgb}{0.00,0.00,0.55}
\definecolor {dodgerblue2}         {rgb}{0.11,0.53,0.93}
\definecolor {dodgerblue3}         {rgb}{0.09,0.45,0.80}
\definecolor {dodgerblue4}         {rgb}{0.06,0.31,0.55}
\definecolor {steelblue1}          {rgb}{0.39,0.72,1.00}
\definecolor {steelblue2}          {rgb}{0.36,0.67,0.93}
\definecolor {steelblue3}          {rgb}{0.31,0.58,0.80}
\definecolor {steelblue4}          {rgb}{0.21,0.39,0.55}
\definecolor {deepskyblue2}        {rgb}{0.00,0.70,0.93}
\definecolor {deepskyblue3}        {rgb}{0.00,0.60,0.80}
\definecolor {deepskyblue4}        {rgb}{0.00,0.41,0.55}
\definecolor {skyblue1}            {rgb}{0.53,0.81,1.00}
\definecolor {skyblue2}            {rgb}{0.49,0.75,0.93}
\definecolor {skyblue3}            {rgb}{0.42,0.65,0.80}
\definecolor {skyblue4}            {rgb}{0.29,0.44,0.55}
\definecolor {lightskyblue1}       {rgb}{0.69,0.89,1.00}
\definecolor {lightskyblue2}       {rgb}{0.64,0.83,0.93}
\definecolor {lightskyblue3}       {rgb}{0.55,0.71,0.80}
\definecolor {lightskyblue4}       {rgb}{0.38,0.48,0.55}
\definecolor {slategray1}          {rgb}{0.78,0.89,1.00}
\definecolor {slategray2}          {rgb}{0.73,0.83,0.93}
\definecolor {slategray3}          {rgb}{0.62,0.71,0.80}
\definecolor {slategray4}          {rgb}{0.42,0.48,0.55}
\definecolor {lightsteelblue1}     {rgb}{0.79,0.88,1.00}
\definecolor {lightsteelblue2}     {rgb}{0.74,0.82,0.93}
\definecolor {lightsteelblue3}     {rgb}{0.64,0.71,0.80}
\definecolor {lightsteelblue4}     {rgb}{0.43,0.48,0.55}
\definecolor {lightblue1}          {rgb}{0.75,0.94,1.00}
\definecolor {lightblue2}          {rgb}{0.70,0.87,0.93}
\definecolor {lightblue3}          {rgb}{0.60,0.75,0.80}
\definecolor {lightblue4}          {rgb}{0.41,0.51,0.55}
\definecolor {lightcyan2}          {rgb}{0.82,0.93,0.93}
\definecolor {lightcyan3}          {rgb}{0.71,0.80,0.80}
\definecolor {lightcyan4}          {rgb}{0.48,0.55,0.55}
\definecolor {paleturquoise1}      {rgb}{0.73,1.00,1.00}
\definecolor {paleturquoise2}      {rgb}{0.68,0.93,0.93}
\definecolor {paleturquoise3}      {rgb}{0.59,0.80,0.80}
\definecolor {paleturquoise4}      {rgb}{0.40,0.55,0.55}
\definecolor {cadetblue1}          {rgb}{0.60,0.96,1.00}
\definecolor {cadetblue2}          {rgb}{0.56,0.90,0.93}
\definecolor {cadetblue3}          {rgb}{0.48,0.77,0.80}
\definecolor {cadetblue4}          {rgb}{0.33,0.53,0.55}
\definecolor {turquoise1}          {rgb}{0.00,0.96,1.00}
\definecolor {turquoise2}          {rgb}{0.00,0.90,0.93}
\definecolor {turquoise3}          {rgb}{0.00,0.77,0.80}
\definecolor {turquoise4}          {rgb}{0.00,0.53,0.55}
\definecolor {cyan2}               {rgb}{0.00,0.93,0.93}
\definecolor {cyan3}               {rgb}{0.00,0.80,0.80}
\definecolor {cyan4}               {rgb}{0.00,0.55,0.55}
\definecolor {darkslategray1}      {rgb}{0.59,1.00,1.00}
\definecolor {darkslategray2}      {rgb}{0.55,0.93,0.93}
\definecolor {darkslategray3}      {rgb}{0.47,0.80,0.80}
\definecolor {darkslategray4}      {rgb}{0.32,0.55,0.55}
\definecolor {aquamarine2}         {rgb}{0.46,0.93,0.78}
\definecolor {aquamarine4}         {rgb}{0.27,0.55,0.45}
\definecolor {darkseagreen1}       {rgb}{0.76,1.00,0.76}
\definecolor {darkseagreen2}       {rgb}{0.71,0.93,0.71}
\definecolor {darkseagreen3}       {rgb}{0.61,0.80,0.61}
\definecolor {darkseagreen4}       {rgb}{0.41,0.55,0.41}
\definecolor {seagreen1}           {rgb}{0.33,1.00,0.62}
\definecolor {seagreen2}           {rgb}{0.31,0.93,0.58}
\definecolor {seagreen3}           {rgb}{0.26,0.80,0.50}
\definecolor {palegreen1}          {rgb}{0.60,1.00,0.60}
\definecolor {palegreen2}          {rgb}{0.56,0.93,0.56}
\definecolor {palegreen3}          {rgb}{0.49,0.80,0.49}
\definecolor {palegreen4}          {rgb}{0.33,0.55,0.33}
\definecolor {springgreen2}        {rgb}{0.00,0.93,0.46}
\definecolor {springgreen3}        {rgb}{0.00,0.80,0.40}
\definecolor {springgreen4}        {rgb}{0.00,0.55,0.27}
\definecolor {green2}              {rgb}{0.00,0.93,0.00}
\definecolor {green3}              {rgb}{0.00,0.80,0.00}
\definecolor {green4}              {rgb}{0.00,0.55,0.00}
\definecolor {chartreuse2}         {rgb}{0.46,0.93,0.00}
\definecolor {chartreuse3}         {rgb}{0.40,0.80,0.00}
\definecolor {chartreuse4}         {rgb}{0.27,0.55,0.00}
\definecolor {olivedrab1}          {rgb}{0.75,1.00,0.24}
\definecolor {olivedrab2}          {rgb}{0.70,0.93,0.23}
\definecolor {olivedrab4}          {rgb}{0.41,0.55,0.13}
\definecolor {darkolivegreen1}     {rgb}{0.79,1.00,0.44}
\definecolor {darkolivegreen2}     {rgb}{0.74,0.93,0.41}
\definecolor {darkolivegreen3}     {rgb}{0.64,0.80,0.35}
\definecolor {darkolivegreen4}     {rgb}{0.43,0.55,0.24}
\definecolor {khaki1}              {rgb}{1.00,0.96,0.56}
\definecolor {khaki2}              {rgb}{0.93,0.90,0.52}
\definecolor {khaki3}              {rgb}{0.80,0.78,0.45}
\definecolor {khaki4}              {rgb}{0.55,0.53,0.31}
\definecolor {lightgoldenrod1}     {rgb}{1.00,0.93,0.55}
\definecolor {lightgoldenrod2}     {rgb}{0.93,0.86,0.51}
\definecolor {lightgoldenrod3}     {rgb}{0.80,0.75,0.44}
\definecolor {lightgoldenrod4}     {rgb}{0.55,0.51,0.30}
\definecolor {lightyellow2}        {rgb}{0.93,0.93,0.82}
\definecolor {lightyellow3}        {rgb}{0.80,0.80,0.71}
\definecolor {lightyellow4}        {rgb}{0.55,0.55,0.48}
\definecolor {yellow2}             {rgb}{0.93,0.93,0.00}
\definecolor {yellow3}             {rgb}{0.80,0.80,0.00}
\definecolor {yellow4}             {rgb}{0.55,0.55,0.00}
\definecolor {gold2}               {rgb}{0.93,0.79,0.00}
\definecolor {gold3}               {rgb}{0.80,0.68,0.00}
\definecolor {gold4}               {rgb}{0.55,0.46,0.00}
\definecolor {goldenrod1}          {rgb}{1.00,0.76,0.15}
\definecolor {goldenrod2}          {rgb}{0.93,0.71,0.13}
\definecolor {goldenrod3}          {rgb}{0.80,0.61,0.11}
\definecolor {goldenrod4}          {rgb}{0.55,0.41,0.08}
\definecolor {darkgoldenrod1}      {rgb}{1.00,0.73,0.06}
\definecolor {darkgoldenrod2}      {rgb}{0.93,0.68,0.05}
\definecolor {darkgoldenrod3}      {rgb}{0.80,0.58,0.05}
\definecolor {darkgoldenrod4}      {rgb}{0.55,0.40,0.03}
\definecolor {rosybrown1}          {rgb}{1.00,0.76,0.76}
\definecolor {rosybrown2}          {rgb}{0.93,0.71,0.71}
\definecolor {rosybrown3}          {rgb}{0.80,0.61,0.61}
\definecolor {rosybrown4}          {rgb}{0.55,0.41,0.41}
\definecolor {indianred1}          {rgb}{1.00,0.42,0.42}
\definecolor {indianred2}          {rgb}{0.93,0.39,0.39}
\definecolor {indianred3}          {rgb}{0.80,0.33,0.33}
\definecolor {indianred4}          {rgb}{0.55,0.23,0.23}
\definecolor {sienna1}             {rgb}{1.00,0.51,0.28}
\definecolor {sienna2}             {rgb}{0.93,0.47,0.26}
\definecolor {sienna3}             {rgb}{0.80,0.41,0.22}
\definecolor {sienna4}             {rgb}{0.55,0.28,0.15}
\definecolor {burlywood1}          {rgb}{1.00,0.83,0.61}
\definecolor {burlywood2}          {rgb}{0.93,0.77,0.57}
\definecolor {burlywood3}          {rgb}{0.80,0.67,0.49}
\definecolor {burlywood4}          {rgb}{0.55,0.45,0.33}
\definecolor {wheat1}              {rgb}{1.00,0.91,0.73}
\definecolor {wheat2}              {rgb}{0.93,0.85,0.68}
\definecolor {wheat3}              {rgb}{0.80,0.73,0.59}
\definecolor {wheat4}              {rgb}{0.55,0.49,0.40}
\definecolor {tan1}                {rgb}{1.00,0.65,0.31}
\definecolor {tan2}                {rgb}{0.93,0.60,0.29}
\definecolor {tan4}                {rgb}{0.55,0.35,0.17}
\definecolor {chocolate1}          {rgb}{1.00,0.50,0.14}
\definecolor {chocolate2}          {rgb}{0.93,0.46,0.13}
\definecolor {chocolate3}          {rgb}{0.80,0.40,0.11}
\definecolor {firebrick1}          {rgb}{1.00,0.19,0.19}
\definecolor {firebrick2}          {rgb}{0.93,0.17,0.17}
\definecolor {firebrick3}          {rgb}{0.80,0.15,0.15}
\definecolor {firebrick4}          {rgb}{0.55,0.10,0.10}
\definecolor {brown1}              {rgb}{1.00,0.25,0.25}
\definecolor {brown2}              {rgb}{0.93,0.23,0.23}
\definecolor {brown3}              {rgb}{0.80,0.20,0.20}
\definecolor {brown4}              {rgb}{0.55,0.14,0.14}
\definecolor {salmon1}             {rgb}{1.00,0.55,0.41}
\definecolor {salmon2}             {rgb}{0.93,0.51,0.38}
\definecolor {salmon3}             {rgb}{0.80,0.44,0.33}
\definecolor {salmon4}             {rgb}{0.55,0.30,0.22}
\definecolor {lightsalmon2}        {rgb}{0.93,0.58,0.45}
\definecolor {lightsalmon3}        {rgb}{0.80,0.51,0.38}
\definecolor {lightsalmon4}        {rgb}{0.55,0.34,0.26}
\definecolor {orange2}             {rgb}{0.93,0.60,0.00}
\definecolor {orange3}             {rgb}{0.80,0.52,0.00}
\definecolor {orange4}             {rgb}{0.55,0.35,0.00}
\definecolor {darkorange1}         {rgb}{1.00,0.50,0.00}
\definecolor {darkorange2}         {rgb}{0.93,0.46,0.00}
\definecolor {darkorange3}         {rgb}{0.80,0.40,0.00}
\definecolor {darkorange4}         {rgb}{0.55,0.27,0.00}
\definecolor {coral1}              {rgb}{1.00,0.45,0.34}
\definecolor {coral2}              {rgb}{0.93,0.42,0.31}
\definecolor {coral3}              {rgb}{0.80,0.36,0.27}
\definecolor {coral4}              {rgb}{0.55,0.24,0.18}
\definecolor {tomato2}             {rgb}{0.93,0.36,0.26}
\definecolor {tomato3}             {rgb}{0.80,0.31,0.22}
\definecolor {tomato4}             {rgb}{0.55,0.21,0.15}
\definecolor {orangered2}          {rgb}{0.93,0.25,0.00}
\definecolor {orangered3}          {rgb}{0.80,0.22,0.00}
\definecolor {orangered4}          {rgb}{0.55,0.15,0.00}
\definecolor {red2}                {rgb}{0.93,0.00,0.00}
\definecolor {red3}                {rgb}{0.80,0.00,0.00}
\definecolor {red4}                {rgb}{0.55,0.00,0.00}
\definecolor {deeppink2}           {rgb}{0.93,0.07,0.54}
\definecolor {deeppink3}           {rgb}{0.80,0.06,0.46}
\definecolor {deeppink4}           {rgb}{0.55,0.04,0.31}
\definecolor {hotpink1}            {rgb}{1.00,0.43,0.71}
\definecolor {hotpink2}            {rgb}{0.93,0.42,0.65}
\definecolor {hotpink3}            {rgb}{0.80,0.38,0.56}
\definecolor {hotpink4}            {rgb}{0.55,0.23,0.38}
\definecolor {pink1}               {rgb}{1.00,0.71,0.77}
\definecolor {pink2}               {rgb}{0.93,0.66,0.72}
\definecolor {pink3}               {rgb}{0.80,0.57,0.62}
\definecolor {pink4}               {rgb}{0.55,0.39,0.42}
\definecolor {lightpink1}          {rgb}{1.00,0.68,0.73}
\definecolor {lightpink2}          {rgb}{0.93,0.64,0.68}
\definecolor {lightpink3}          {rgb}{0.80,0.55,0.58}
\definecolor {lightpink4}          {rgb}{0.55,0.37,0.40}
\definecolor {palevioletred1}      {rgb}{1.00,0.51,0.67}
\definecolor {palevioletred2}      {rgb}{0.93,0.47,0.62}
\definecolor {palevioletred3}      {rgb}{0.80,0.41,0.54}
\definecolor {palevioletred4}      {rgb}{0.55,0.28,0.36}
\definecolor {maroon1}             {rgb}{1.00,0.20,0.70}
\definecolor {maroon2}             {rgb}{0.93,0.19,0.65}
\definecolor {maroon3}             {rgb}{0.80,0.16,0.56}
\definecolor {maroon4}             {rgb}{0.55,0.11,0.38}
\definecolor {violetred1}          {rgb}{1.00,0.24,0.59}
\definecolor {violetred2}          {rgb}{0.93,0.23,0.55}
\definecolor {violetred3}          {rgb}{0.80,0.20,0.47}
\definecolor {violetred4}          {rgb}{0.55,0.13,0.32}
\definecolor {magenta2}            {rgb}{0.93,0.00,0.93}
\definecolor {magenta3}            {rgb}{0.80,0.00,0.80}
\definecolor {magenta4}            {rgb}{0.55,0.00,0.55}
\definecolor {orchid1}             {rgb}{1.00,0.51,0.98}
\definecolor {orchid2}             {rgb}{0.93,0.48,0.91}
\definecolor {orchid3}             {rgb}{0.80,0.41,0.79}
\definecolor {orchid4}             {rgb}{0.55,0.28,0.54}
\definecolor {plum1}               {rgb}{1.00,0.73,1.00}
\definecolor {plum2}               {rgb}{0.93,0.68,0.93}
\definecolor {plum3}               {rgb}{0.80,0.59,0.80}
\definecolor {plum4}               {rgb}{0.55,0.40,0.55}
\definecolor {mediumorchid1}       {rgb}{0.88,0.40,1.00}
\definecolor {mediumorchid2}       {rgb}{0.82,0.37,0.93}
\definecolor {mediumorchid3}       {rgb}{0.71,0.32,0.80}
\definecolor {mediumorchid4}       {rgb}{0.48,0.22,0.55}
\definecolor {darkorchid1}         {rgb}{0.75,0.24,1.00}
\definecolor {darkorchid2}         {rgb}{0.70,0.23,0.93}
\definecolor {darkorchid3}         {rgb}{0.60,0.20,0.80}
\definecolor {darkorchid4}         {rgb}{0.41,0.13,0.55}
\definecolor {purple1}             {rgb}{0.61,0.19,1.00}
\definecolor {purple2}             {rgb}{0.57,0.17,0.93}
\definecolor {purple3}             {rgb}{0.49,0.15,0.80}
\definecolor {purple4}             {rgb}{0.33,0.10,0.55}
\definecolor {mediumpurple1}       {rgb}{0.67,0.51,1.00}
\definecolor {mediumpurple2}       {rgb}{0.62,0.47,0.93}
\definecolor {mediumpurple3}       {rgb}{0.54,0.41,0.80}
\definecolor {mediumpurple4}       {rgb}{0.36,0.28,0.55}
\definecolor {thistle1}            {rgb}{1.00,0.88,1.00}
\definecolor {thistle2}            {rgb}{0.93,0.82,0.93}
\definecolor {thistle3}            {rgb}{0.80,0.71,0.80}
\definecolor {thistle4}            {rgb}{0.55,0.48,0.55}
\definecolor {gray1}               {rgb}{0.01,0.01,0.01}
\definecolor {gray2}               {rgb}{0.02,0.02,0.02}
\definecolor {gray3}               {rgb}{0.03,0.03,0.03}
\definecolor {gray4}               {rgb}{0.04,0.04,0.04}
\definecolor {gray5}               {rgb}{0.05,0.05,0.05}
\definecolor {gray6}               {rgb}{0.06,0.06,0.06}
\definecolor {gray7}               {rgb}{0.07,0.07,0.07}
\definecolor {gray8}               {rgb}{0.08,0.08,0.08}
\definecolor {gray9}               {rgb}{0.09,0.09,0.09}
\definecolor {gray10}              {rgb}{0.10,0.10,0.10}
\definecolor {gray11}              {rgb}{0.11,0.11,0.11}
\definecolor {gray12}              {rgb}{0.12,0.12,0.12}
\definecolor {gray13}              {rgb}{0.13,0.13,0.13}
\definecolor {gray14}              {rgb}{0.14,0.14,0.14}
\definecolor {gray15}              {rgb}{0.15,0.15,0.15}
\definecolor {gray16}              {rgb}{0.16,0.16,0.16}
\definecolor {gray17}              {rgb}{0.17,0.17,0.17}
\definecolor {gray18}              {rgb}{0.18,0.18,0.18}
\definecolor {gray19}              {rgb}{0.19,0.19,0.19}
\definecolor {gray20}              {rgb}{0.20,0.20,0.20}
\definecolor {gray21}              {rgb}{0.21,0.21,0.21}
\definecolor {gray22}              {rgb}{0.22,0.22,0.22}
\definecolor {gray23}              {rgb}{0.23,0.23,0.23}
\definecolor {gray24}              {rgb}{0.24,0.24,0.24}
\definecolor {gray25}              {rgb}{0.25,0.25,0.25}
\definecolor {gray26}              {rgb}{0.26,0.26,0.26}
\definecolor {gray27}              {rgb}{0.27,0.27,0.27}
\definecolor {gray28}              {rgb}{0.28,0.28,0.28}
\definecolor {gray29}              {rgb}{0.29,0.29,0.29}
\definecolor {gray30}              {rgb}{0.30,0.30,0.30}
\definecolor {gray31}              {rgb}{0.31,0.31,0.31}
\definecolor {gray32}              {rgb}{0.32,0.32,0.32}
\definecolor {gray33}              {rgb}{0.33,0.33,0.33}
\definecolor {gray34}              {rgb}{0.34,0.34,0.34}
\definecolor {gray35}              {rgb}{0.35,0.35,0.35}
\definecolor {gray36}              {rgb}{0.36,0.36,0.36}
\definecolor {gray37}              {rgb}{0.37,0.37,0.37}
\definecolor {gray38}              {rgb}{0.38,0.38,0.38}
\definecolor {gray39}              {rgb}{0.39,0.39,0.39}
\definecolor {gray40}              {rgb}{0.40,0.40,0.40}
\definecolor {gray42}              {rgb}{0.42,0.42,0.42}
\definecolor {gray43}              {rgb}{0.43,0.43,0.43}
\definecolor {gray44}              {rgb}{0.44,0.44,0.44}
\definecolor {gray45}              {rgb}{0.45,0.45,0.45}
\definecolor {gray46}              {rgb}{0.46,0.46,0.46}
\definecolor {gray47}              {rgb}{0.47,0.47,0.47}
\definecolor {gray48}              {rgb}{0.48,0.48,0.48}
\definecolor {gray49}              {rgb}{0.49,0.49,0.49}
\definecolor {gray50}              {rgb}{0.50,0.50,0.50}
\definecolor {gray51}              {rgb}{0.51,0.51,0.51}
\definecolor {gray52}              {rgb}{0.52,0.52,0.52}
\definecolor {gray53}              {rgb}{0.53,0.53,0.53}
\definecolor {gray54}              {rgb}{0.54,0.54,0.54}
\definecolor {gray55}              {rgb}{0.55,0.55,0.55}
\definecolor {gray56}              {rgb}{0.56,0.56,0.56}
\definecolor {gray57}              {rgb}{0.57,0.57,0.57}
\definecolor {gray58}              {rgb}{0.58,0.58,0.58}
\definecolor {gray59}              {rgb}{0.59,0.59,0.59}
\definecolor {gray60}              {rgb}{0.60,0.60,0.60}
\definecolor {gray61}              {rgb}{0.61,0.61,0.61}
\definecolor {gray62}              {rgb}{0.62,0.62,0.62}
\definecolor {gray63}              {rgb}{0.63,0.63,0.63}
\definecolor {gray64}              {rgb}{0.64,0.64,0.64}
\definecolor {gray65}              {rgb}{0.65,0.65,0.65}
\definecolor {gray66}              {rgb}{0.66,0.66,0.66}
\definecolor {gray67}              {rgb}{0.67,0.67,0.67}
\definecolor {gray68}              {rgb}{0.68,0.68,0.68}
\definecolor {gray69}              {rgb}{0.69,0.69,0.69}
\definecolor {gray70}              {rgb}{0.70,0.70,0.70}
\definecolor {gray71}              {rgb}{0.71,0.71,0.71}
\definecolor {gray72}              {rgb}{0.72,0.72,0.72}
\definecolor {gray73}              {rgb}{0.73,0.73,0.73}
\definecolor {gray74}              {rgb}{0.74,0.74,0.74}
\definecolor {gray75}              {rgb}{0.75,0.75,0.75}
\definecolor {gray76}              {rgb}{0.76,0.76,0.76}
\definecolor {gray77}              {rgb}{0.77,0.77,0.77}
\definecolor {gray78}              {rgb}{0.78,0.78,0.78}
\definecolor {gray79}              {rgb}{0.79,0.79,0.79}
\definecolor {gray80}              {rgb}{0.80,0.80,0.80}
\definecolor {gray81}              {rgb}{0.81,0.81,0.81}
\definecolor {gray82}              {rgb}{0.82,0.82,0.82}
\definecolor {gray83}              {rgb}{0.83,0.83,0.83}
\definecolor {gray84}              {rgb}{0.84,0.84,0.84}
\definecolor {gray85}              {rgb}{0.85,0.85,0.85}
\definecolor {gray86}              {rgb}{0.86,0.86,0.86}
\definecolor {gray87}              {rgb}{0.87,0.87,0.87}
\definecolor {gray88}              {rgb}{0.88,0.88,0.88}
\definecolor {gray89}              {rgb}{0.89,0.89,0.89}
\definecolor {gray90}              {rgb}{0.90,0.90,0.90}
\definecolor {gray91}              {rgb}{0.91,0.91,0.91}
\definecolor {gray92}              {rgb}{0.92,0.92,0.92}
\definecolor {gray93}              {rgb}{0.93,0.93,0.93}
\definecolor {gray94}              {rgb}{0.94,0.94,0.94}
\definecolor {gray95}              {rgb}{0.95,0.95,0.95}
\definecolor {gray97}              {rgb}{0.97,0.97,0.97}
\definecolor {gray98}              {rgb}{0.98,0.98,0.98}
\definecolor {gray99}              {rgb}{0.99,0.99,0.99}
\definecolor {darkgrey}            {rgb}{0.66,0.66,0.66}

\newcommand{\TODO}[1]{{}}

\newcommand{\ignore}[1]{}
\newcommand{\ignoreinshort}[1]{}
 
\def\makenewenumerate#1#2{%
\newcounter{cnt#1}
\newenvironment{#1}%
{\begin{list}{\makebox[0pt][r]{#2}}%
{\setlength{\itemsep}{0pt}% 
 \setlength{\parsep}{.2em}%
 \setlength{\leftmargin}{1.5em}%
 \setlength{\labelwidth}{.4em}%
 \usecounter{cnt#1}}}
{\end{list}}}

\makenewenumerate{myenumerate}{\arabic{cntmyenumerate}.}

\makenewenumerate{renumerate}{\rm(\roman{cntrenumerate})}
\makenewenumerate{renumerateprime}{\rm(\roman{cntrenumerateprime}$'$)}
\makenewenumerate{renumeratesecond}{\rm(\roman{cntrenumeratesecond}$''$)}

\makenewenumerate{aenumerate}{({\it\alph{cntaenumerate}})}
\makenewenumerate{aenumerateprime}{({\it\alph{cntaenumerateprime}$'$})}
\makenewenumerate{aenumeratesecond}{({\it\alph{cntaenumeratesecond}$''$})}

\def\newplaintheorem#1#2{%
\newtheorem{#1plain}{#2}% %% RS: mon mi piace l'indice di sezione
\newenvironment{#1}{\begin{#1plain}\rm }{\end{#1plain}}}

\newplaintheorem{notation}{Notation}
\newplaintheorem{cexample}{Counter-Example}

\newcommand{\pos}{\phantom{\neg}}

\newcommand{\solver}[1]{{\sc TabularAllSAT{#1}}\xspace}

\newcommand\mysout{\bgroup \markoverwith{{-}}\ULon}
\newcommand\nosout{\bgroup \markoverwith{{ }}\ULon}
\definecolor{mygray}{rgb}{0.90,0.90,0.90}
\definecolor{mywhite}{rgb}{1.00,1.00,1.00}

\newcommand{\ti}[1]{\ensuremath{\sf{t}^{(#1)}}\xspace}
\newcommand{\tn}[1]{\ti{n}}

\newcommand{\GMCHANGE}[1]{\textcolor{black}{#1}}
\newcommand{\GSCHANGE}[1]{\textcolor{black}{#1}}
\newcommand{\GSCHANGEBIS}[1]{\textcolor{black}{#1}}

\newcommand{\OptFN}[2]{{\ifx&#2&\ensuremath{#1}\else\ensuremath{#1(#2)}\fi}}

\newcommand{\exdone}{\ensuremath{\hfill\diamond}}
\newcommand{\trail}{\emph{T}}
\newcommand{\simplify}{{\sc Check-Literal}}

\newtheorem{definition}{Definition}
\newtheorem{example}{Example}

\begin{document}

\maketitle

\begin{abstract}
A basic algorithm for enumerating disjoint propositional models (disjoint AllSAT) is based on adding blocking clauses incrementally, ruling out previously found models. On the one hand, blocking clauses have the potential to reduce the number of generated models exponentially, as they can handle partial models. On the other hand, the introduction of a large number of blocking clauses affects memory consumption and drastically slows down unit propagation. 
 We propose a new approach that allows for enumerating disjoint partial models with no need for blocking clauses by integrating: Conflict-Driven Clause-Learning (CDCL), Chronological Backtracking (CB), and methods for shrinking models (Implicant Shrinking). Experiments clearly show the benefits of our novel approach.
\end{abstract}

\section{Introduction}
\label{sec:introduction}

All-Solution Satisfiability Problem (AllSAT) is an extension of SAT that requires finding all possible solutions of a propositional formula. AllSAT has been heavily applied in the field of hardware and software verification. For instance, AllSAT can be used to generate test suites for programs automatically \cite{khurshid2004case} and for bounded and unbounded model checking \cite{jin2005efficient}.
\GMCHANGE{
Recently AllSAT has found applications in artificial intelligence. For example, \cite{spallitta2022smt} exploits AllSMT (a variant of AllSAT dealing with first-order logic theories) for probabilistic inference in hybrid domains. AllSAT has also been applied to data mining to deal with the frequent itemset mining problem \cite{dlala2016comparative}. }
\GSCHANGE{
Lastly, model counting over first-order logic theories ($\#$SMT) \cite{chistikov2015approximate} relies on AllSAT too.
}

Exploring the complete search space efficiently is a major concern in AllSAT. For a formula $F$ with $n$ variables, there are $2^n$ possible total assignments. Generating all of these assignments explicitly would require exponential space complexity. To mitigate the issue, we can use \textit{partial} models to obtain compact representations of a set of solutions. If a partial model does not explicitly assign the truth value of a variable, then it means that its truth value does not impact the satisfiability of that assignment, thus two assignments are represented by the partial one. In problems with $n$ variables, a partial assignment with $m$ variables covers $2^{n - m}$ total assignments in one shot. 

\GSCHANGE{
The literature distinguishes between enumeration with repetitions (AllSAT) and enumeration without repetitions (\textit{disjoint AllSAT}). Whereas covering the same model may not be problematic for certain applications (e.g. predicate abstraction \cite{lahiri2003symbolic}), it can result in an incorrect final solution for other contexts, such as Weighted Model Integration \cite{morettin2019advanced} and $\#$SMT \cite{chistikov2015approximate}. In this paper, we will address disjoint AllSAT.}

\GMCHANGE{SAT-based propositional enumeration algorithms} can be grouped into two main categories: blocking solvers, and non-blocking solvers.

\textit{Blocking} AllSAT solvers \cite{mcmillan2002applying,jin2005efficient, yu2014all} rely on Conflict Driven Clause-Learning (CDCL) and non-chronological
backtracking (NCB) to return the set of all satisfying assignments. 
\GMCHANGE{They work by repeatedly adding blocking clauses to the formula after each model is found, which rules out the previous set of satisfying assignments until all possible satisfying assignments have been found. These blocking clauses ensure that the solver does not return the same satisfying assignment multiple times and that the search space is efficiently scanned \cite{1562927}.} Although blocking solvers are straightforward to implement and can be adapted
to retrieve partial assignments, they become inefficient when the input formula $F$ has a high number of models,
as an \GSCHANGE{\textbf{exponential number}} of blocking clauses might be added to make sure the entire search space is visited. As the number of blocking clauses increases, unit propagation becomes more difficult, resulting in degraded performance. 

\textit{Non-blocking} AllSAT solvers \cite{grumberg2004memory, li2004novel} overcome this issue by not introducing blocking clauses and
by implementing \textit{chronological backtracking} (CB) \cite{nadel2018chronological}: after a conflict arises, they backtrack on the search tree by
updating the most recently instantiated variable. Chronological backtracking guarantees not to cover the same model of a formula
multiple times without the typical CPU-time blow-up caused by blocking clauses. \GSCHANGE{The major drawback of this class of AllSAT solvers is that they only generate \textbf{total assignments}}. Moreover, regions of the search space with no solution cannot be escaped easily.

\cite{mohle2019combining} proposes a new formal calculus of a disjunctive model counting algorithm combining the best
features of chronological backtracking and CDCL, but without providing an implementation or experimental results.
 \GMCHANGE{In \cite{sebastiani2020you, mohle2020four, mohle2021enumerating} the authors discuss the calculus behind different approaches to determine if a partial assignment satisfies a formula when chronological backtracking is implemented in the CDCL procedure. However, both works rely on dual reasoning, which could perform badly when a high number of variables is involved (SAT and QBF oracle calls required by \cite{mohle2020four} may be expensive).}

\paragraph*{Contributions}

In this work, we propose a novel AllSAT procedure to perform disjoint partial enumeration of propositional formulae by combining the
best of current AllSAT state-of-the-art literature: ($i$) CDCL, to escape search branches where no satisfiable assignments can be
found; ($ii$) chronological backtracking, to ensure no blocking clauses are introduced; ($iii$) efficient implicant shrinking, to reduce
in size partial assignments, by exploiting the 2-literal watching scheme.
We have implemented the aforementioned ideas in a tool that we refer to as \solver{} and compared its performance against other publicly available state-of-the-art AllSAT tools using a variety of benchmarks, including both crafted and SATLIB instances. Our experimental results show that \solver{} outperforms all other solvers on nearly all benchmarks, 
demonstrating the benefits of our approach.

\section{Background}
\label{sec:preliminaries}

\subsection{Notation} 

 We assume $F$ is a propositional formula defined on the set of Boolean variables $V = \{v_1, ..., v_n\}$, with cardinality $|V|$. A \textit{literal} $\ell$ is a variable $v$ or its negation $\neg v$. $L(V)$ denotes the set of literals on V. We implicitly remove double negations: if $\ell$ is $\neg v$, by $\neg \ell$ we mean $v$ rather than $\neg\neg v$. A \textit{clause} is the disjunction of literals $\bigvee_{\ell\in c} \ell$. A \textit{cube} is the conjunction of literals $\bigwedge_{\ell\in c} \ell$. 

 The function $M: V \mapsto \{\top, \perp \}$ mapping variables in $F$ to their truth value is known as \textit{assignment}. An assignment can be represented by either a set of literals $\{\ell_1, ..., \ell_n\}$ or a cube conjoining all literals in the assignment $\ell_1 \wedge ... \wedge \ell_n$.
 We distinguish between \textit{total assignments} $\eta$ or \textit{partial assignments} $\mu$ depending on whether all variables are mapped to a truth value or not, respectively.
 
 A \textit{trail} is an ordered sequence of literals $I = \ell_1, ..., \ell_n$ with no duplicate variables. The empty trail is represented by $\varepsilon$. Two trails can be conjoined one after the other $I = KL$, assuming $K$ and $L$ have no variables in common. We use superscripts to mark literals in a trail $I$: $\ell^d$ indicates a literal assigned during the decision phase, whereas $\ell^*$ refers to literals whose truth value is negated due to chronological backtracking after finding a model (we will refer to this action as \textit{flipping}). Trails can be seen as ordered \emph{total} (resp. \emph{partial}) assignments; for the sake
of simplicity, we will refer to them as \emph{total} (resp. \emph{partial}) trails.

 \begin{definition}
     The \emph{decision level function} $\delta(V) \mapsto \mathbb{N} \cup \{\infty\}$ returns the decision level of variable $V$, where $\infty$ means unassigned. We extend this concept to literals ($\delta(\ell) = \delta(V(\ell))$) and clauses ($\delta(C) = \{max(\delta(\ell)) |\ell \in C\}$).
 \end{definition}

  \begin{definition}
     The \emph{decision literal function} $\sigma(dl) \mapsto L(V) \cup \{\varepsilon\}$ returns the decision literal of level $dl$. If we have not decided on a literal at level $dl$ yet, we return $\varepsilon$. 
 \end{definition}

\begin{definition}
    The \emph{reason function} $\rho(\ell)$ returns the reason that forced literal $\ell$ to be assigned a truth value:
\begin{itemize}
    \item \textsc{DECISION}, if the literal is assigned by the decision selection procedure;
    \item \textsc{UNIT}, if the literal is unit propagated at decision level 0, thus it is an initial literal;
    \item \textsc{PROPAGATED($c$)}, if the literal is unit propagated at a decision level higher than 0 due to clause $c$;
\end{itemize} 
\end{definition}

 \subsection{The 2-Watched Literal Scheme}
 \label{sec:2watched}

 The \emph{2-watched literal scheme} \cite{moskewicz2001chaff} is an
indexing technique that efficiently checks if the currently-assigned
literals do not cause a conflict. For every clause, two literals are tracked. If at least one of the two literals is set to $\top$, then the clause is satisfied. If one of the two literals is set to $\perp$, then we scan the clause searching for a new literal $\ell'$ that can be paired with the remaining one, being sure $\ell'$ is not mapped to $\perp$. If we reach the end of the clause \GSCHANGE{and both watches for that clause are set to false}, then we know the current assignment falsifies the formula. The 2-watched literal scheme is implemented through watch lists.

 \begin{definition}
The \emph{watch list function} $\omega(\ell)$ returns the set of clauses $\{c_1,...,c_n\}$ currently watched by literal $\ell$.
\end{definition}

 \subsection{CDCL and Non-chronological Backtracking}

%\textcolor{blue}{TODO: rewrite it}
 Conflict Driven Clause Learning (CDCL) is the most popular SAT-solving technique \cite{marques1999grasp}. It is an extension of the older Davis-Putnam-Logemann-Loveland (DPLL) algorithm \cite{davis1962machine}, improving the latter by dynamically learning new clauses during the search process and using them to drive backtracking.

Every time the current trail falsifies a formula $F$, the SAT solver generates a conflict clause $c$ starting from the falsified clause, by repeatedly resolving against the clauses which caused unit propagation of falsified literals. This clause is then learned by the solver and added to $F$. Depending on $c$, we backtrack to flip the value of one literal, potentially jumping more than one decision level (thus we talk about \emph{non-chronological backtracking}, or NBC). CDCL and non-chronological backtracking allow for escaping regions of the search space where no satisfying assignments are admitted, which benefits both SAT and AllSAT solving.
The idea behind conflict clauses has been extended in AllSAT to learn clauses from partial satisfying assignments (known in the literature as \emph{good learning} or \emph{blocking clauses} \cite{bayardo2000counting,morgado2005good}) to ensure no total assignment is covered twice.

 \subsection{Chronological Backtracking}

%\textcolor{blue}{TODO: rewrite it} 
\ignore{
\GMCHANGE{
The idea of exploiting chronological backtracking (CB) was first presented in GRASP \cite{marques1999grasp} in the context of clause learning, and then revamped for both SAT and AllSAT in \cite{nadel2018chronological, mohle2019backing}.} 
}
\GSCHANGEBIS{Chronological backtracking (CB) is the core of the original DPLL algorithm. Considered inefficient for SAT solving once NBC was presented in \cite{moskewicz2001chaff}, it was recently revamped for both SAT and AllSAT in \cite{nadel2018chronological, mohle2019backing}.}
The intuition is that non-chronological backtracking after conflict analysis can lead to redundant work, due to some assignments that could be repeated later on during the search.
Instead, independent of the generated conflict clause $c$ we chronologically backtrack and flip the last decision literal in the trail. Consequently, we explore the search space systematically and efficiently, ensuring no assignment is covered twice during execution. Chronological backtracking combined with CDCL is effective in SAT solving when dealing with satisfiable instances. In AllSAT solving, it guarantees blocking clauses are no more needed to ensure termination.

\section{Enumerating Disjoint Partial Models without Blocking Clauses}
\label{sec:solver}

 We propose a novel approach that allows for enumerating disjoint partial models with no need for blocking clauses, by
 integrating: Conflict-Driven Clause-Learning ({\bf CDCL}), to escape search branches where no satisfiable assignments can be found;
 Chronological Backtracking ({\bf CB}), to ensure no blocking clauses are introduced; and methods for shrinking models ({\bf Implicant Shrinking}),
 to reduce in size partial assignments, by exploiting the 2-watched literal schema.
 
 To this extent, \cite{mohle2019combining} discusses a formal calculus to combine CDCL and CB for propositional model counting, strongly related to the task we want to achieve. We take the calculus presented in that paper as the theoretical foundation on top of which we build our algorithms, and refer to that paper for more details.

\subsection{Disjoint AllSAT by Integrating CDCL and CB}
\label{sec:solver-search}

The work in \cite{mohle2019combining} exclusively describes the calculus and a formal proof of correctness for a model counting
framework on top of CDCL and CB,  with neither any algorithm nor any reference in modern state-of-the-art solvers. To this extent, we
start by presenting an AllSAT procedure for the search algorithm combining the two techniques, which are reported in this section. \GMCHANGE{In particular, we highlight the major differences to a classical AllSAT solver implemented on top of CDCL and NBC}.

\GMCHANGE{Algorithm \ref{algo:chronocdcl} presents the main search loop of the AllSAT algorithm. The goal is to find a total trail \trail{} that satisfies $F$. At each decision level, it iteratively decides one of the unassigned variables in $F$ and assigns a truth value (lines \ref{algo:decide}-\ref{algo:decide-end}); it then performs unit propagation
(line \ref{algo:unit}) until either a conflict is reached (lines \ref{algo:conflict}-\ref{algo:conflict-end}), or no other variable can be
unit propagated leading to a satisfying total assignment (lines \ref{algo:noconflict}-\ref{algo:noconflict-end}) or \textsc{Decide} has to be called again (lines \ref{algo:decide}-\ref{algo:decide-end}).}

\begin{algorithm}[t]
\begin{algorithmic}[1]
  \caption[A]{{\sc Chrono-CDCL}($F, V$)}% 
  \label{algo:chronocdcl}
  \STATE $T \leftarrow \varepsilon$
  \STATE $dl \leftarrow 0$
  \WHILE{\textbf{true}}
  \STATE $T, c \leftarrow$ \textsc{UnitPropagation()} \label{algo:unit}
  \IF{$c \neq \varepsilon$}\label{algo:conflict} 
    \STATE \underline{\textsc{AnalyzeConflict}($T,c, dl$)} \label{algo:conflict-stop}
    \ELSIF{$|T| = |V|$} \label{algo:noconflict}
        \STATE \underline{\textsc{AnalyzeAssignment($T, dl$)}} \label{algo:noconflict-end}
  \ELSE
    \STATE \textsc{Decide}($T$) \label{algo:decide}
    \STATE $dl \leftarrow dl + 1$ \label{algo:decide-end}
  \ENDIF
  \ENDWHILE
\end{algorithmic}
\end{algorithm}

\GMCHANGE{Notice that the main loop is identical to an AllSAT solver based on non-chronological CDCL; the only differences are embedded in the procedure to get the conflict and the partial assignments. (From now on, we underline the lines that differ from the baseline CDCL AllSAT solver.)}

\begin{algorithm}[t]
\begin{algorithmic}[1]
  \caption[A]{{\sc AnalyzeConflict}($T, c, dl$)} \label{algo:get-conflict-full}
  \IF{\underline{$\delta(c) < dl$}} \label{algo:backfirst}
        \STATE \underline{$T \leftarrow$ \textsc{Backtrack($\delta(c)$)}} \label{algo:backfirst-end}
    \ENDIF
    \IF{\textcolor{black}{$dl = 0$}} \label{algo:terminate}
        \STATE \textcolor{black}{\textbf{terminate with all models found}} \label{algo:terminate-end}
    \ENDIF
    \STATE $\langle uip, c' \rangle \leftarrow$ \textsc{LastUIP-Analysis()} \label{algo:conflictanalysis}
    \STATE \underline{$T \leftarrow$ \textsc{Backtrack($dl-1$)}} \label{algo:conflict-start}
    \STATE $T.push(\neg uip)$
    \STATE $\rho(\neg uip) \leftarrow$ \textsc{Propagated($c'$)} \label{algo:conflict-end}
\end{algorithmic}
\end{algorithm}

\GMCHANGE{Suppose {\sc UnitPropagation} finds a conflict, returning one clause $c$ in $F$ which is falsified by the current trail \trail{}, so that we invoke {\sc AnalyzeConflict}. Algorithm \ref{algo:get-conflict-full} shows the procedure to either generate the conflict clause or stop the search for new assignments if all models have been found.} 

We first compute the maximum assignment level of all literals in the conflicting clause $c$ and
    backtrack to that decision level (lines \ref{algo:backfirst}-\ref{algo:backfirst-end}) if strictly smaller than $dl$. This \GMCHANGE{additional step, not contemplated by AllSAT solvers that use NCB,} is necessary to support out-of-order
    assignments, the core insight in chronological backtracking when integrated into CDCL as described in \cite{nadel2018chronological}. 
    
    \GMCHANGE{Apart from this first step, Algorithm \ref{algo:get-conflict-full} behaves similarly to a standard conflict analysis algorithm.} If the solver
    reaches decision level 0 at this point, it means there are no more variables to flip and the whole search space has been visited,
    and we can terminate the algorithm (lines \ref{algo:terminate}-\ref{algo:terminate-end}). Otherwise, we perform conflict analysis up to the last Unique Implication Point (last UIP\GSCHANGEBIS{, i.e. the decision variable at the current decision level}), retrieving the conflict clause $c'$ (line \ref{algo:conflictanalysis}),
    as proposed in~\cite{mohle2019combining}. Finally, we perform backtracking \GMCHANGE{(notice how we force chronological backtracking independently from the decision level of the conflict clause)}, push the flipped UIP into the trail, and set $c'$ as its assignment reason for the flipping (lines \ref{algo:conflict-start}-\ref{algo:conflict-end}).  

    \begin{algorithm}[t]
\begin{algorithmic}[1]
  \caption[A]{{\sc AnalyzeAssignment}($T, dl$)} \label{algo:get-assignment-full}
  \STATE \underline{$dl' \leftarrow $ \textsc{Implicant-Shrinking($T$)}} \label{algo:implishrink}
        \IF{\underline{$dl' < dl$}} \label{algo:backshrink}
            \STATE \underline{$T \leftarrow$ \textsc{Backtrack($dl'$)}} \label{algo:backshrink-end}
        \ENDIF
        \STATE \textbf{store model} $T$ \label{algo:print}
        \IF{$dl' = 0$} \label{algo:exit-true}
            \STATE \textbf{terminate with all models found} \label{algo:exit-true-end}
        \ELSE
            \STATE $\ell_{flip} \leftarrow \neg (\sigma(dl'))$ \label{algo:chronoback}
            \STATE \underline{$T \leftarrow$ \textsc{Backtrack($dl' - 1$)}}
            \STATE $T.push(\ell_{flip})$
            \STATE $\rho(\ell_{flip}) =$ \textsc{Backtrue} \label{algo:chronoback-end}
        \ENDIF
    
\end{algorithmic}
\end{algorithm}

Suppose instead that every variable is assigned a truth value without generating conflicts (Algorithm \ref{algo:chronocdcl}, line \ref{algo:noconflict}); \GMCHANGE{then 
the current total trail \trail{} satisfies $F$, and we invoke {\sc AnalyzeAssignment}. Algorithm \ref{algo:get-assignment-full} shows the steps to possibly shrink the assignment, store it and continue the search.}

First, \textsc{Implicant-Shrinking} checks if, for some decision level $dl'$, we can backtrack up to $dl' < dl$ and obtain a partial trail still satisfying the formula (Algorithm \ref{algo:get-assignment-full}, lines \ref{algo:implishrink}-\ref{algo:backshrink-end}). 
(We discuss the details of chronological implicant shrinking in the next subsection.)
We can produce the current assignment from the current trail \trail{} (line \ref{algo:print}). Then we check if all variables in \trail{} are assigned at decision level $0$.
If this is the case, then this means that we found the last assignment to cover $F$, so that we can end the search (lines \ref{algo:exit-true}-\ref{algo:exit-true-end}). Otherwise, we perform chronological backtracking, flipping the truth value of the currently highest decision variables and searching for a new total trail \trail{} satisfying $F$ (lines \ref{algo:chronoback}-\ref{algo:chronoback-end}).

We remark that in
\cite{mohle2019combining} it is implicitly assumed that one can determine if a partial trail satisfies the formula right after being generated, whereas modern SAT
solvers cannot check this fact efficiently, and
detect satisfaction only when trails are total. To cope with this issue, in our approach the
partial trail satisfying the formula is computed {\em a posteriori} from the total one by implicant shrinking. \GSCHANGEBIS{Moreover, the mutual exclusivity among different assignments is guaranteed, since the shrinking of the assignments is performed so that the generated partial assignments fall under the conditions of Section 3 in \cite{mohle2019combining}).}   

Notice that the calculus discussed in \cite{mohle2019combining} assumes the last UIP is the termination criteria for the conflict analysis. We provide the following counter-example to
show that the first UIP does not guarantee mutual exclusivity between returned assignments.

\begin{example}
\label{ex:fuzz}
    Let $F$ be the propositional formula:
    $$ F = \overbrace{(x_1 \vee \neg x_2)}^{c_1} \wedge \overbrace{(x_1 \vee \neg x_3)}^{c_2} \wedge \overbrace{(\neg x_1 \vee \neg x_2)}^{c_3} $$
    For the sake of simplicity, we assume \textsc{Chrono-CDCL} to return total truth assignments. If the initial variable ordering
    is $x_3, x_2, x_1$ (all set to false) then the first two total and the third partial trails generated by Algorithm \ref{algo:chronocdcl} are: 
    \begin{equation*}
        T_1 = \textcolor{black}{\neg x_3^d \neg x_2^d \neg x_1^d};\ \ 
    T_2 = \textcolor{black}{\neg x_3^d \neg x_2^d x_1^*};\ \ 
    T_3 = \textcolor{black}{\neg x_3^d x_2^*}
    \end{equation*}
    Notice how $T_3$ leads to a falsifying assignment: $x_2$ forces $x_1$ due to $c_1$ and $\neg x_1$ due to $c_3$ at the same time. A
    conflict arises and we adopt the first UIP algorithm to stop conflict analysis. We identify $x_2$ as the first unique implication point (UIP) and construct the conflict clause $\neg x_2$. Since this is a unit clause, we force its negation $\neg x_2$ as an initial unit. We can now set $x_3$ and $x_1$ to $\perp$ and obtain a satisfying assignment. The resulting total trail $T = \neg x_3 \neg x_2 \neg x_1$ is covered \textbf{twice} during the search process.
    \exdone{}
    
\end{example}

\GMCHANGE{
We also emphasize that the incorporation of restarts in the search algorithm (or any method that implicitly exploits restarts, such as rephasing) is not feasible, as reported in \cite{mohle2019combining}.
}

\subsection{Chronological Implicant Shrinking}

\label{sec:partial}

\begin{algorithm}[t]
\begin{algorithmic}[1]
  \caption[A]{{\sc Implicant-Shrinking}($T$)}% 
  \label{algo:shrinking}
  \STATE $b \leftarrow 0$ 
  \STATE $T' \leftarrow T$
  \WHILE{$T' \neq \varepsilon$} \label{algo:while-lift}
      \STATE $\ell \leftarrow T'.pop()$ \label{algo:etapop}
      \IF{$\rho(\ell) \neq $ \textsc{DECISION}} \label{algo:decisionnot}
            \STATE $b \leftarrow max(b, \delta(\ell))$ \label{algo:decisionnot-lineend}
                
    \ELSIF{$\delta(\ell) > b$}\label{algo:compare}
        \STATE $b \leftarrow$ \textsc{Check-Literal}$(\ell, b, T')$ \label{algo:simp}
    \ELSIF{$\delta(\ell) = 0$ or ($\delta(\ell) = b$ and $\rho(\ell) =$  \textsc{Decision})} \label{alg:shrinking-ending}
        \STATE \textbf{break} \label{alg:shrinking-ending-end}
    \ENDIF
       
  \ENDWHILE
  \STATE \textbf{return} $b$
\end{algorithmic}
\end{algorithm}

\noindent Effectively shrinking a total trail \trail{} when chronological backtracking is enabled is not trivial. 

In principle, we could add a flag for each clause $c$ stating if $c$ is currently satisfied by the partial assignment or not, and check the status of all flags iteratively adding literals to the trail. Despite being easy to integrate into an AllSAT solver and avoiding assigning all variables a truth value, this approach is unfeasible in practice: every time a new literal $\ell$ is added/removed from the trail, we should check and eventually update the value of the flags of clauses containing it. In the long term, this would negatively affect performances, particularly when the formula has a large number of models.

Also, relying on implicant shrinking algorithms from the literature for NCB-based AllSAT solvers does not work for chronological backtracking. Prime-implicant shrinking algorithms do not guarantee the mutual exclusivity between different assignments, so that they are not useful in the context of disjoint AllSAT. Other assignment-minimization algorithms, as in \cite{toda2016implementing}, work under the assumption that a blocking clause is introduced. 

For instance, suppose we perform disjoint AllSAT on the formula $F = x_1 \vee x_2$ and the ordered trail is $T_1 = x_1^d x_2^d$. A general assignment minimization algorithm could retrieve the partial assignment $\mu = x_2$ satisfying $F$, but obtaining it by using chronological backtracking is not possible (it would require us to remove $x_1$ from the trail despite being assigned at a lower decision level than $x_2$) unless blocking clauses are introduced. 

In this context, we need an implicant shrinking algorithm such that: ($i$) it is compatible with chronological backtracking, i.e. we remove variables assigned at level $dl$ or higher as if they have never been assigned; ($ii$) it tries to cut the highest amount of literals while still ensuring mutual exclusivity.

 Considering all the aforementioned issues, we propose a {\it chronological implicant shrinking} algorithm that uses state-of-the-art SAT solver data structures (thus without requiring dual encoding), which is described in Algorithm \ref{algo:shrinking}. 

The idea is to pick literals from the current trail starting from the latest assigned literals (lines \ref{algo:while-lift}-\ref{algo:etapop}) and determine the lowest decision level $b$ to backtrack and shrink the implicant. First, we check if $\ell$ was not assigned by \textsc{Decide} (line \ref{algo:decisionnot}). If this is the case, we set $b$ to be at least as high as the decision level of $\ell$ ($\delta(\ell)$), ensuring that it will not be dropped by implicant shrinking (line \ref{algo:decisionnot-lineend}), since $\ell$ has a role in performing disjoint AllSAT.

If this is not the case, we compare its decision level $\delta(\ell)$ to $b$ (line \ref{algo:compare}). If $\delta(\ell) > b$, then
we actively check if it is necessary for \trail{} to satisfy $F$ (line \ref{algo:simp}) and set $b$ accordingly. Two versions of \simplify{}  will be presented.

If $\ell$ is either an initial literal (i.e. assigned at decision level 0) or both $\rho(\ell) =$ \textsc{Decision} and $\delta(\ell) = b$ hold, all literals in the trail assigned before $\ell$ would have a decision level lower or equal than $b$. This means that we can exit the loop early (lines \ref{alg:shrinking-ending}-\ref{alg:shrinking-ending-end}), since scanning further the trail would be unnecessary.
Finally, if none of the above conditions holds, we can assume that $b$ is already greater than $\delta(\ell)$, and we can move on to the next literal in the trail.

\subsubsection{Checking Literals Using 2-Watched Lists.}
\label{sec:implicant-watches}

In \cite{deharbe2013computing} the authors propose an algorithm to shorten total assignments and obtain a prime implicant by using watch lists. We adopted the ideas from this work and adapted them to be integrated into CB-based AllSAT solving, which we present in Algorithm \ref{algo:dynamic}.  

\begin{algorithm}[t]
\begin{algorithmic}[1]
    \caption[A]{\simplify($\ell$, $b, T'$)}% 
    \label{algo:dynamic}
    \FOR{$ c \in \omega(\ell)$} \label{algo:dynamic-for1}
        \IF{$\exists \ell' \in c$ s.t. $\ell' \neq \ell$ and $\ell' \in T'$ } \label{algo:dynamic-for2}
    %\STATE $\omega() = \omega(\ell') + \langle\ell_\omega, c\rangle$ \label{algo:dynamic-body}
    %\STATE $\omega(\ell) = \omega(\ell) - \langle\ell_\omega, c\rangle$
    %\STATE $\omega(\ell_\omega) = \omega(\ell_\omega) - \langle\ell, c\rangle + \langle\ell', c\rangle$
            \STATE Watch $c$ by $\ell'$ instead of $\ell$\label{algo:dynamic-body}
            \ELSE
            \STATE $b \leftarrow max(b, \delta(\ell))$ 
        \ENDIF
    \label{algo:dynamic-update}
    \ENDFOR
    \RETURN $b$
    \label{algo:fullwatch-end}
\end{algorithmic}
\end{algorithm}

For each literal $\ell$ we check its watch list $\omega(\ell)$ (line \ref{algo:dynamic-for1}). For each clause $c$ in $\omega(\ell)$ we are interested in finding a literal $\ell'$ such that: ($i$) $\ell'$ is not $\ell$ itself, ($ii$) $\ell'$ satisfies $c$ and it is in the current trail $T'$ so that it has not already been checked by \textsc{Implicant-Shrinking} (line \ref{algo:dynamic-for2}). If it exists, we update the watch lists, so that now $\ell'$ watches $c$ instead of $\ell$, then we move on to the next clause (line \ref{algo:dynamic-body}). If no replacement for $\ell$ is available, then $\ell$ is the only remaining literal that guarantees $c$ is satisfied, and we cannot reduce it. We update $b$ accordingly, ensuring $\ell$ would not be minimized by setting $b$ to a value higher or equal than $\delta(\ell)$ (line \ref{algo:dynamic-update}). 
\ignore{We stress the fact that once \textsc{Implicant-Shrinking} terminates, all watch lists should be restored to their value before the procedure was called, otherwise some of the admissible models of $F$ would not be found by the search algorithm.}

\begin{example}
\label{ex:dynamic}
    Let $F$ be the following propositional formula:
\begin{equation*}
      F = \overbrace{(x_1 \vee x_2 \vee x_3)}^{c_1} 
\end{equation*}
     $F$ is satisfied by 7 different total assignments:
\ignore{
    \begin{center}
$\begin{array}{llll} 
            \{\pos \textcolor{black}{x_1,}&\pos \textcolor{black}{x_2,}&\pos \textcolor{black}{x_3}&\},
            \\
            \{ \textcolor{black}{\neg x_1,}&\pos \textcolor{black}{x_2,} &\pos \textcolor{black}{x_3}&\},
            \\
            \{\pos \textcolor{black}{x_1,}&  \textcolor{black}{\neg x_2,} & \pos \textcolor{black}{x_3}&\},
            \\
            \{\textcolor{black}{\neg x_1,}&  \textcolor{black}{\neg x_2,} & \pos \textcolor{black}{x_3}&\},
            
            \\
            \{\pos \textcolor{black}{x_1,}& \pos \textcolor{black}{x_2,} &  \textcolor{black}{\neg x_3}&\},
            
            \\
            \{ \textcolor{black}{\neg x_1,}& \pos \textcolor{black}{x_2,} & \textcolor{black}{\neg x_3}&\},
            \\
            \{\pos \textcolor{black}{x_1,}&  \textcolor{black}{\neg x_2,} &  \textcolor{black}{\neg x_3}&\}
            \\  
        \end{array}$    
\end{center}
}
\begin{equation*}
     \{\pos \textcolor{black}{x_1,}\pos \textcolor{black}{x_2,}\pos \textcolor{black}{x_3}\},
 \{ \textcolor{black}{\neg x_1,}\pos \textcolor{black}{x_2,} \pos \textcolor{black}{x_3}\},            
\{\pos \textcolor{black}{x_1,}  \textcolor{black}{\neg x_2,}  \pos \textcolor{black}{x_3}\},
\end{equation*}
\begin{equation*}
    \{\textcolor{black}{\neg x_1,}  \textcolor{black}{\neg x_2,}  \pos \textcolor{black}{x_3}\},
            \{\pos \textcolor{black}{x_1,} \pos \textcolor{black}{x_2,}   \textcolor{black}{\neg x_3}\},
            \{ \textcolor{black}{\neg x_1,} \pos \textcolor{black}{x_2,}  \textcolor{black}{\neg x_3}\},
\end{equation*}
\begin{equation*}
    \{\pos \textcolor{black}{x_1,}  \textcolor{black}{\neg x_2,}   \textcolor{black}{\neg x_3}\}
\end{equation*}

    When initialized, our solver has the following watch lists:
    \begin{align*}
    &&\omega(x_1) = & \{c_1\};
    &&\omega(x_2) = & \{c_1\};
    &&\omega(x_3) = &\ \emptyset
    \end{align*}
    Algorithm \ref{algo:chronocdcl} can produce the total trail $I_1 = x_3^d x_2^d x_1^d$. \simplify{} starts by minimizing the value of $x_1$. The watch list associated with $x_1$ contains $c_1$, hence we need to substitute $x_1$ with a new literal in clause $c_1$. A suitable substitute exists, namely $x_3$. We update the watch lists according to Algorithm \ref{algo:dynamic}, and obtain: %. The watch lists are now:
    \begin{align*}
    &&\omega(x_1) = &\ \emptyset;
    &&\omega(x_2) = &\{c_1 \};
    &&\omega(x_3) = &\{c_1 \}
    \end{align*}
    Next, \simplify{} eliminates $x_2$ from the current trail: $x_1$ was already cut off, $x_2$ and $x_3$ are the current indexes for $c_1$, and $x_3$ is assigned to $\top$. Since no other variables are available in $c_1$, we must force $x_3$ to be part of the partial assignment, and we set $b$ to 1 to prevent its shrinking. This yields the partial trail $T_1 = {x_3}$.
    
    Chronological backtracking now restores the watched literal indexing to its value before implicant shrinking (in this case the initial state of watch lists) and flips $x_3$ into $\neg x_3$. \textsc{Decide} will then assign $\top$ to both $x_2$ and $x_1$. The new trail $T_2 = \neg x_3^* x_2^d x_1^d$ satisfies $F$. Algorithm \ref{algo:dynamic} drops $x_1$ since $c_1$ is watched by $x_2$ and thus we would still satisfy $F$ without it. $x_2$, on the other hand, is required in $T_2$: $x_3$ is now assigned to $\perp$ and thus cannot substitute $x_2$. We obtain the second partial trail $T_2 = \neg x_3 x_2^d$. Last, we chronologically backtrack and set $x_2$ to $\top$. Being $x_3$ and $x_2$ both $\perp$, \textsc{Unit-Propagation} forces $x_1$ to be $\top$ at level 0. We obtain the last trail satisfying $F$, $T_3 = \neg x_3 \neg x_2 x_1$.

    The final solution is then:

\begin{equation*}
    \{\textcolor{black}{x_3}\}, \{\textcolor{black}{x_2,}  \textcolor{black}{\neg x_3}\}, \{\textcolor{black}{x_1,}  \textcolor{black}{\neg x_2},  \textcolor{black}{\neg x_3}\}
\end{equation*}
\end{example}

\subsubsection{A Faster but Conservative Literal Check.}
\label{sec:lifting}

\iffalse
\begin{algorithm}[t]
\begin{algorithmic}[1]
    \caption[A]{{\sc Simplify-Literal-Static-Watches}($\ell$, $b, I'$)}% 
    \label{algo:static}
    \IF{$\omega(\ell) = \emptyset$} \label{algo:emptywatch}
        \STATE \textbf{return} $b$
    \ENDIF \label{algo:emptywatch-end}
    \FOR{$\ell_\omega \in \omega(\ell)$} \label{algo:fullwatch}
    \IF{$\ell_\omega = \perp$ or $\ell_\omega \notin I$}
    \STATE \textbf{return} $max(b, \delta(\ell))$
    \ELSE
    \STATE $b \leftarrow max(b, \delta(\ell_\omega))$
    \ENDIF
    \ENDFOR \label{algo:fullwatch-end}
    \RETURN $b$
\end{algorithmic}
\end{algorithm}
\fi

\GSCHANGE{In Algorithm \ref{algo:dynamic} the cost of scanning clauses using the 2-watched literal schema during implicant shrinking could result in a bottleneck if plenty of models cover a formula. Bearing this in mind, we propose a lighter variant of Algorithm \ref{algo:dynamic} that does not requires watch lists to be updated.}

Suppose that the current trail $T$ satisfies $F$, which implies that for each clause $c$ in $F$, at least one of the two watched literals of $c$, namely $\ell_1$ and $\ell_2$, is in $T$. 
\GSCHANGE{If \simplify{} tries to remove $\ell_1$ from the trail, instead of checking if there exists another literal in $c$ that satisfies the clause in its place as in line 2 of Algorithm \ref{algo:dynamic}, we simply check the truth value of $\ell_2$ as if the clause $c$ is projected into the binary clause $\ell_1 \vee \ell_2$. If $\ell_2$ is not in $I$, then we force the AllSAT solver to maintain $\ell_1$, setting the backtracking level to at least $\delta(\ell_1)$; otherwise we move on to the next clause watched by it. 
}

It is worth noting that this variant of implicant shrinking is conservative when it comes to dropping literals from the trail. We do not consider the possibility of another literal $\ell'$ watching $c$, is in the current trail $T$, and has a lower decision level than the two literals watching $c$. In such a case, we could set $b$ to $\delta(\ell')$, resulting in a more compact partial assignment. Nonetheless, not scanning the clause can significantly improve performance, making our approach a viable alternative when covering many solutions.

\subsection{Implicit Solution Reasons}
\label{sec:extend-allsat}

Incorporating chronological backtracking into the AllSAT algorithm makes blocking clauses unnecessary. Upon discovering a model, we backtrack chronologically to the most recently assigned decision variable $\ell$ and flip its truth value, as if there were a reason clause $c$ - containing the negated decision literals of \trail{} - that forces the flip. These reason clauses $c$ are typically irrelevant to SAT solving and are not stored in the system. On the other hand, when CDCL is combined with chronological backtracking, \GMCHANGE{these clauses are required for conflict analysis}.

\begin{example}
\label{ex:storing}
    Let $F$ be the same formula from Example \ref{ex:fuzz}.
    We assume the first trail generated by Algorithm \ref{algo:chronocdcl} is $T_1 = \neg x_3^d \neg x_2^d \neg x_1^d$. 
 Algorithm \ref{algo:shrinking} can reduce $x_1$ since $\neg x_2$ suffices to satisfy both $c_1$ and $c_3$. Consequently, we obtain the assignment $\mu_1 = \neg x_3 \wedge \neg x_2$, then flip $\neg x_2$ to $x_2$. The new trail $I_2 = \neg x_3^d x_2^*$ forces $x_1$ to be true due to  $c_1$; then $c_3$ would not be satisfiable anymore and cause the generation of a conflict. The last UIP is $x_3$, so that the reason clause $c'$ forcing $x_2$ to be flipped must be handled by the solver to compute the conflict clause.\exdone{}
\end{example}

To cope with this fact, a straightforward approach would be storing these clauses in memory with no update to the literal watching indexing; this approach would allow for $c$ to be called exclusively by the CDCL procedure without affecting variable propagation. If $F$ admits a large number of models, however, storing these clauses would negatively affect performances, so either we had to frequently call flushing procedures to remove inactive backtrack reason clauses, or we could risk going out of memory to store them.

To overcome the issue, we introduce the notion of \textit{virtual backtrack reason clauses}. When a literal $\ell$ is flipped after a satisfying assignment is found, its reason clause contains the negation of decision literals assigned at a level lower than $\delta(\ell)$ and $\ell$ itself. Consequently, we introduce an additional value, \textsc{Backtrue}, to the possible answers of the reason function $\rho$. This value is used to tag literals flipped after a (possibly partial) assignment is found. When the conflict analysis algorithm encounters a literal $\ell$ having $\rho(\ell) = $ \textsc{Backtrue}, \GMCHANGE{the resolvent can be easily reconstructed by collecting all the decision literals with a lower level than $\ell$ and negating them. This way we do not need to explicitly store these clauses for conflict analysis, allowing us to save time and memory for clause flushing}.

\subsection{Decision Variable Ordering}
\label{sec:ordering}

As shown in \cite{mohle2019combining}, different orders during \textsc{Decide} can lead to a different number of partial trails retrieved if chronological backtracking is enabled. 
After an empirical evaluation, we set {\sf Decide} to select the priority score of a variable depending on the following ordered set of rules.

First, we rely on the Variable State Aware Decaying Sum \textit{(VSADS)} heuristic \cite{huang2005using} and set the priority of a variable according to two weighted factors: ($i$) the count of variable occurrences in the formula, as in the Dynamic Largest Combined Sum (DLCS) heuristics; and ($ii$) an "activity score," which increases when the variable appears in conflict clauses and decreases otherwise, as in the Variable State Independent Decaying Sum (VSIDS) heuristic. If two variables have the same score, we set a higher priority to variables whose watch list is not empty (this is particularly helpful when the lighter variant of the implicant shrinking is used). If there is still a tie, we rely on the lexicographic order of the name of the variables.

\section{Experimental Evaluation}
\label{sec:experiments}

We implemented all the ideas discussed in the paper in a tool we refer to as \solver{}. The code of the algorithm and all benchmarks are available here: \url{https://zenodo.org/records/10397723}. \GMCHANGE{It is built on top of a minimal SAT solver: besides chronological backtracking, it does not have any preprocessing,
restarts and rephasing are disabled, and watching data structures are similar to MiniSAT.}

Experiments are performed on an Intel Xeon Gold 6238R @ 2.20GHz 28 Core machine with 128 GB of RAM, running Ubuntu Linux 20.04. Timeout has been set to 1200 seconds.

\subsection{Benchmarks}

\begin{table*}[t]
\centering
\begin{tabular}{ccccccc}
                    & \solver{} & BDD  & NBC  & MathSAT & BC   & {\sc BC\_Partial} \\
binary clauses (50) & \textbf{30}            & 28   & 21   & 16      & 13   & 18          \\
rnd3sat (410)       & \textbf{410}           & 409  & 396  & 229     & 194  & 210         \\
CSB (1000)          & \textbf{1000}          & \textbf{1000} & \textbf{1000} & 997     & 865  & 636         \\
BMS (500)           & \textbf{499}           & 498  & 498  & 473     & 368  & 353         \\ \hline
Total (1960)        & \textbf{1939}          & 1935 & 1915 & 1715    & 1440 & 1217       
\end{tabular}
\caption{Table reporting the number of instances solved by each solver within the timeout time (1200 seconds).}
\label{tb:table}
\end{table*}

\ignore{
\begin{figure}[t!]
        \centering
        \begin{subfigure}[b]{0.23\textwidth}
            \centering
            \includegraphics[width=0.9\textwidth]{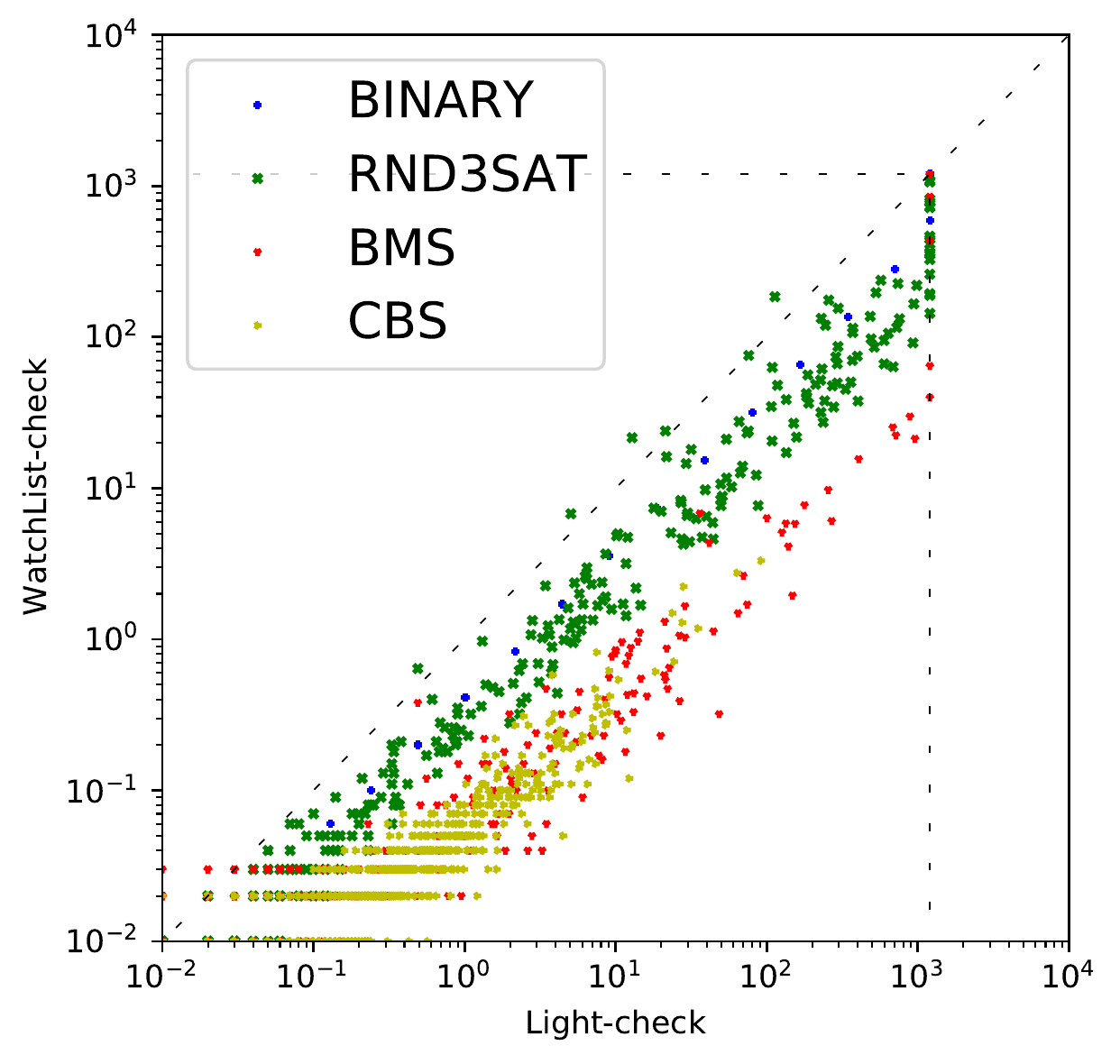}
            \caption%
            {{CPU Time (in seconds)}}    
            \label{fig:binary}
        \end{subfigure}
        \begin{subfigure}[b]{0.23\textwidth}  
            \centering 
            \includegraphics[width=0.9\textwidth]{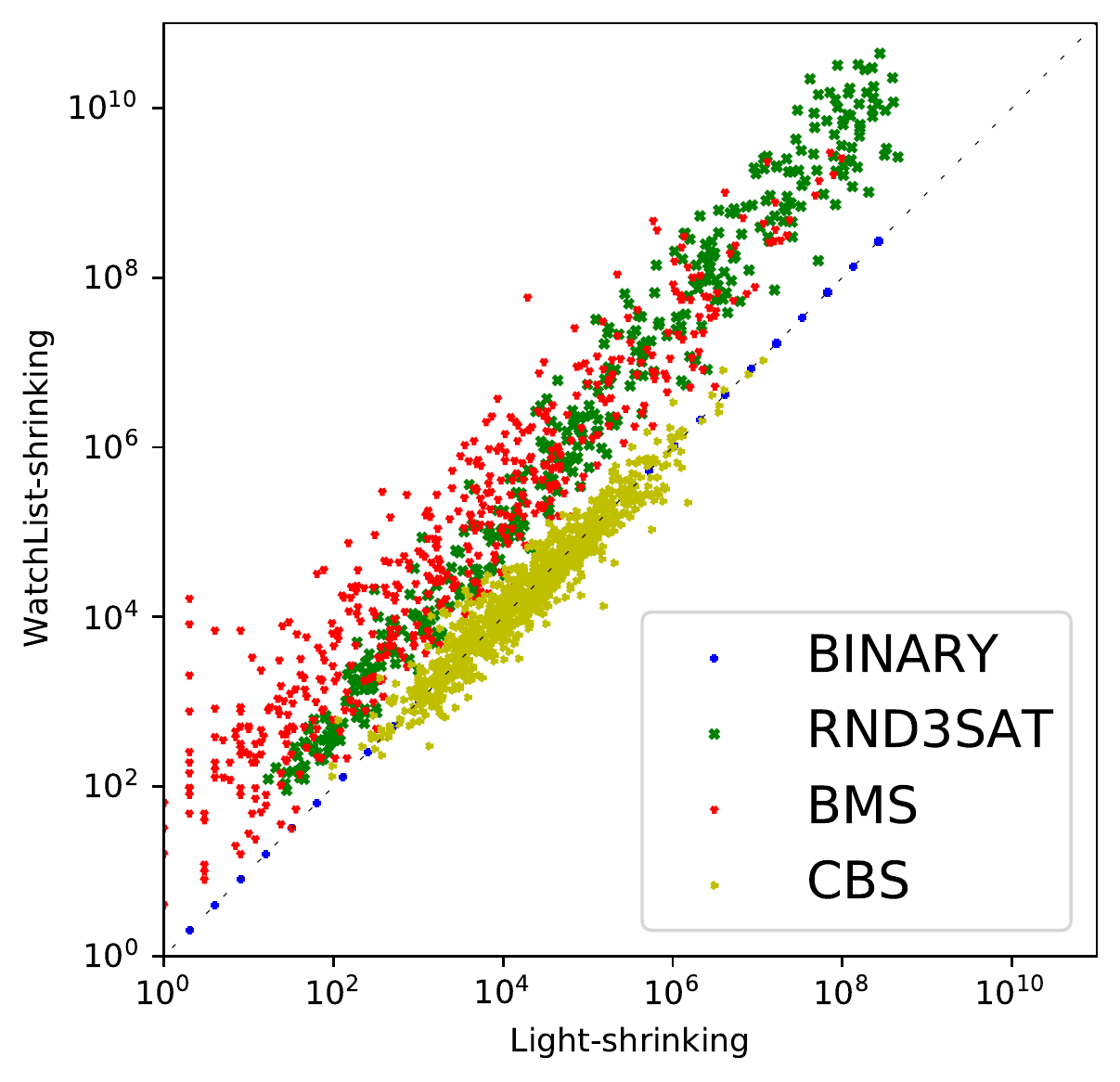}
            \caption[]%
            {{\# of partial models}}    
            \label{fig:pigeon}
        \end{subfigure}
        \caption{Scatter plot comparing CPU time and log-total \# of partial models with the two implicant shrinking algorithms.} 
        \label{fig:scatter-time}
    \end{figure}

}

\begin{figure*}[t!]
        \centering
        \begin{subfigure}[b]{0.34\textwidth}
            \centering
            \includegraphics[width=0.9\textwidth]{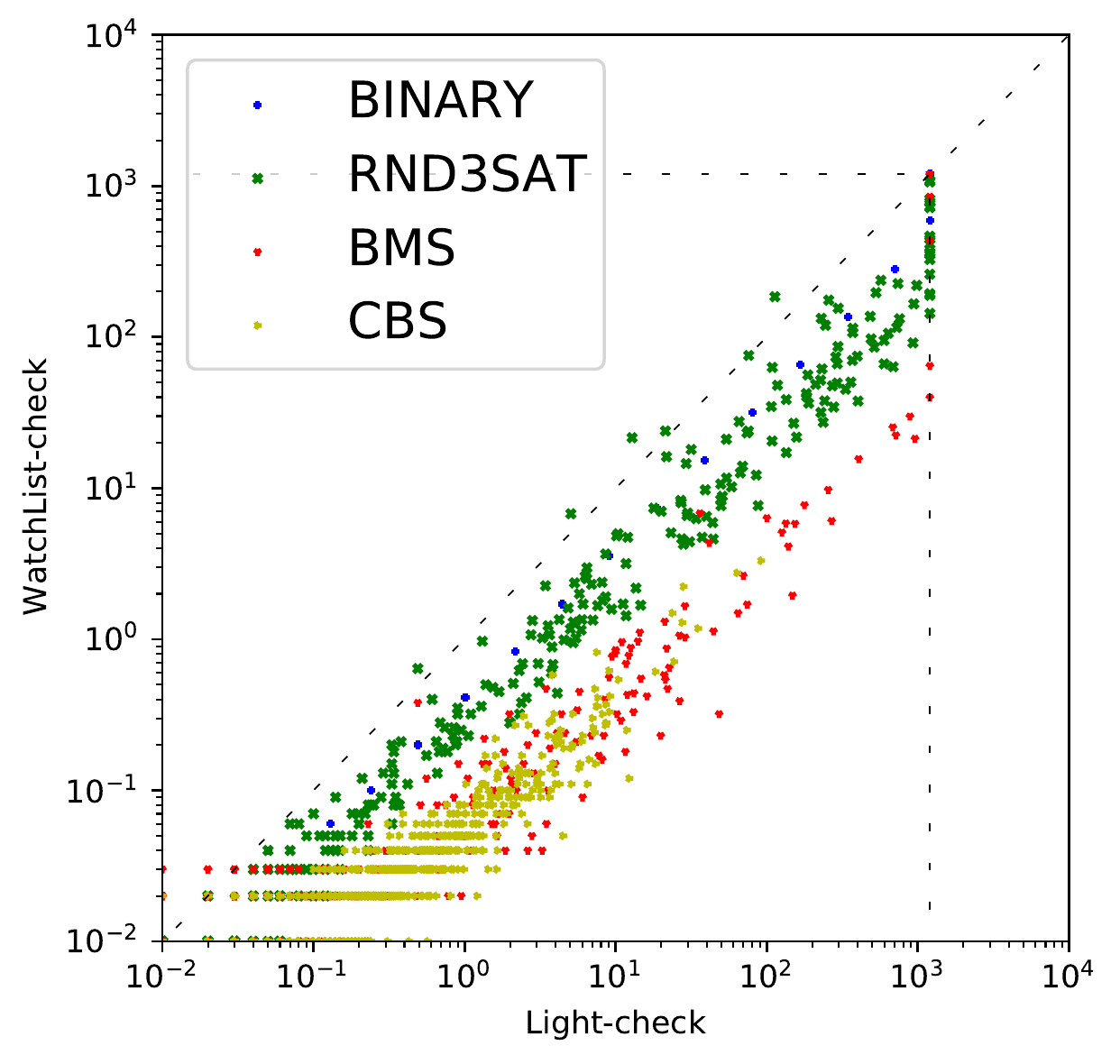}
            \caption%
            {{CPU Time (in seconds)}}    
            \label{fig:binary}
        \end{subfigure}
        \begin{subfigure}[b]{0.34\textwidth}  
            \centering 
            \includegraphics[width=0.9\textwidth]{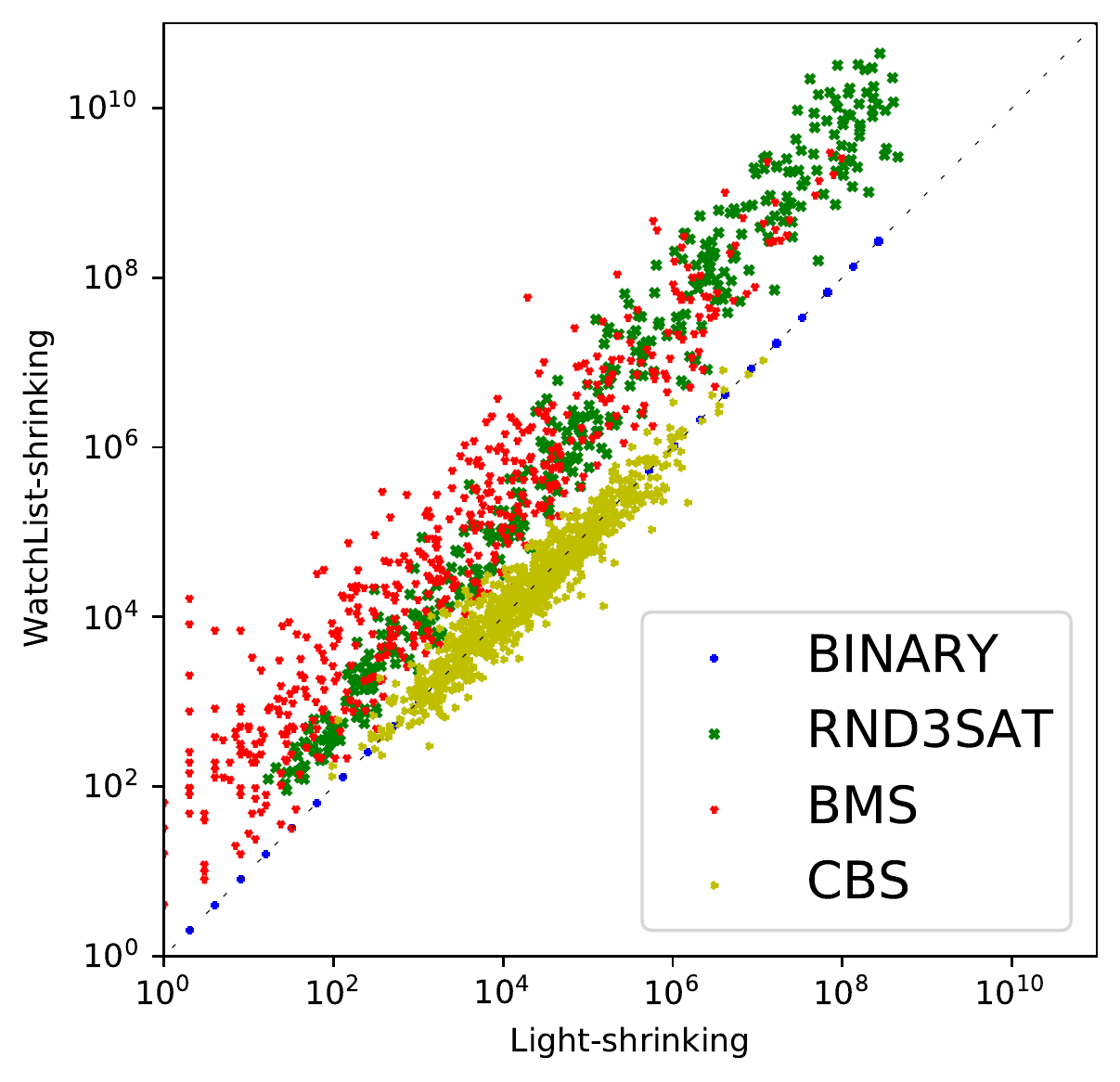}
            \caption[]%
            {{\# of partial models}}    
            \label{fig:pigeon}
        \end{subfigure}
        \caption{Scatter plot comparing CPU time and log-total \# of partial models with the two implicant shrinking algorithms.} 
        \label{fig:scatter-time}
    \end{figure*}

% ORIGINAL SIZE
\ignore{
\begin{figure*}[!t]
    \centering
    \begin{subfigure}[t]{0.19\textwidth}
        \centering
        \includegraphics[width=\textwidth]{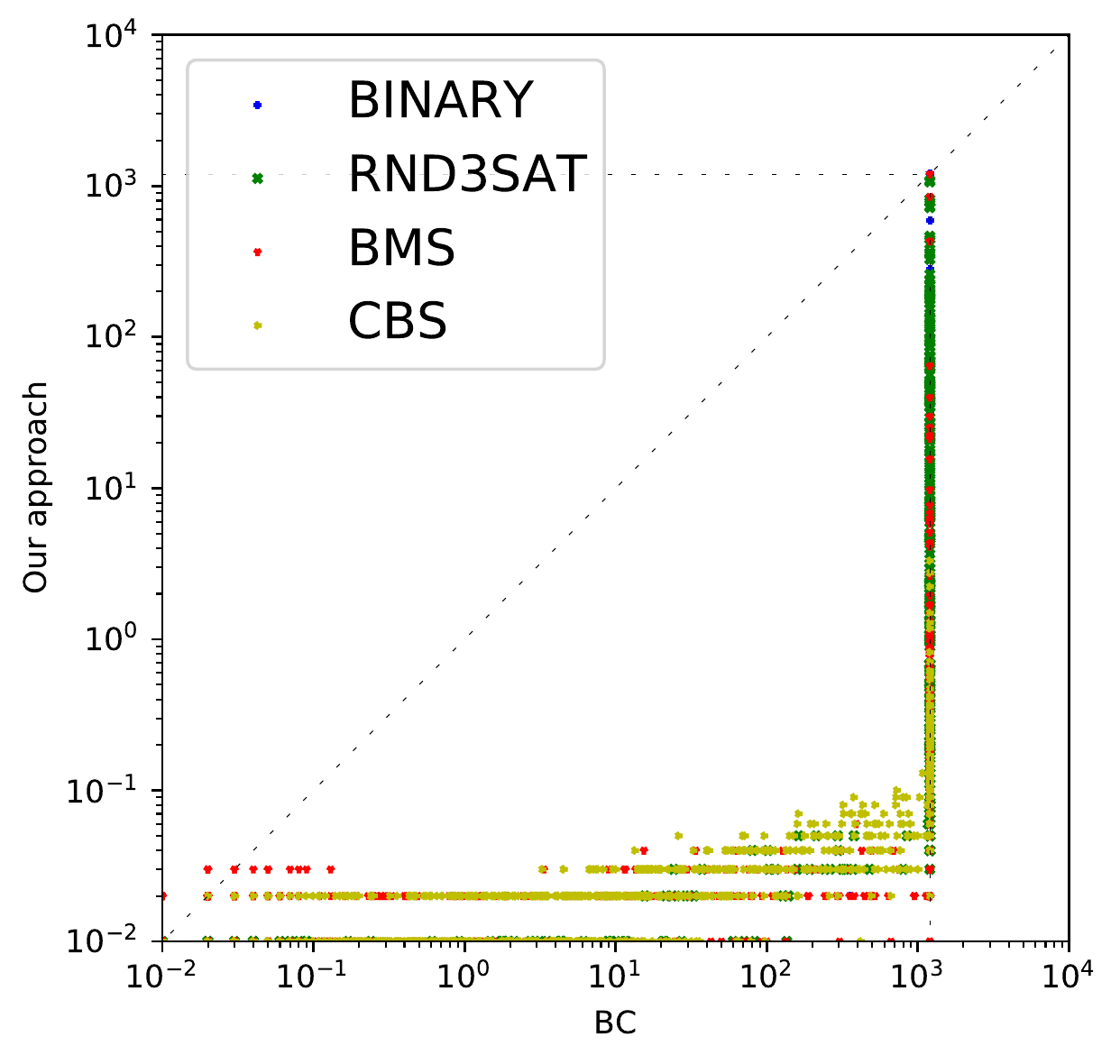}
        \caption%
        {{BC}}    
        \label{fig:soa-binary}
    \end{subfigure}
    \begin{subfigure}[t]{0.19\textwidth}  
        \centering 
        \includegraphics[width=\textwidth]{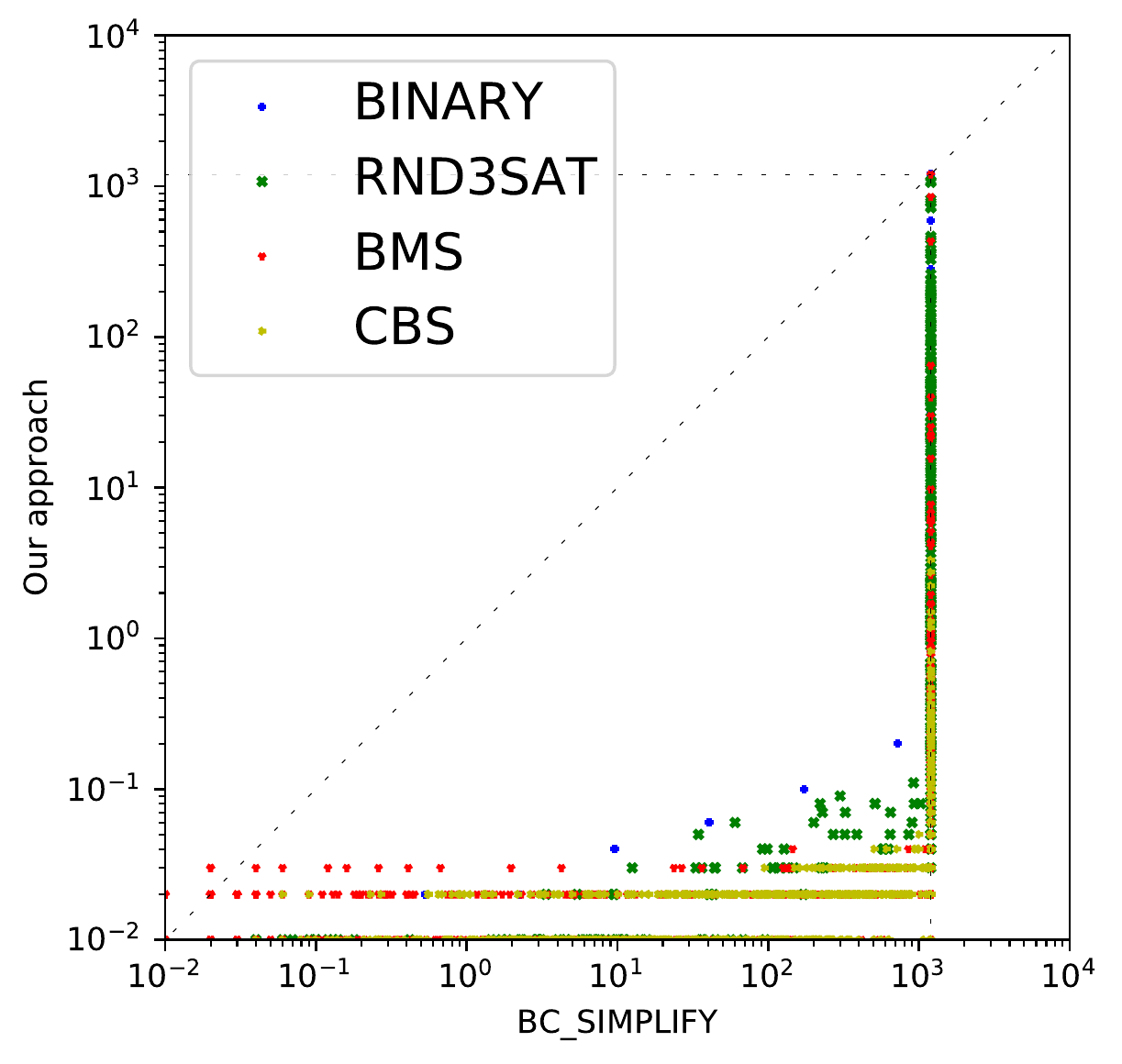}
        \caption[]%
        {{{\sc BC\_Partial}}}    
        \label{fig:soa-CSB}
    \end{subfigure}
    \begin{subfigure}[t]{0.19\textwidth}   
        \centering 
        \includegraphics[width=\textwidth]{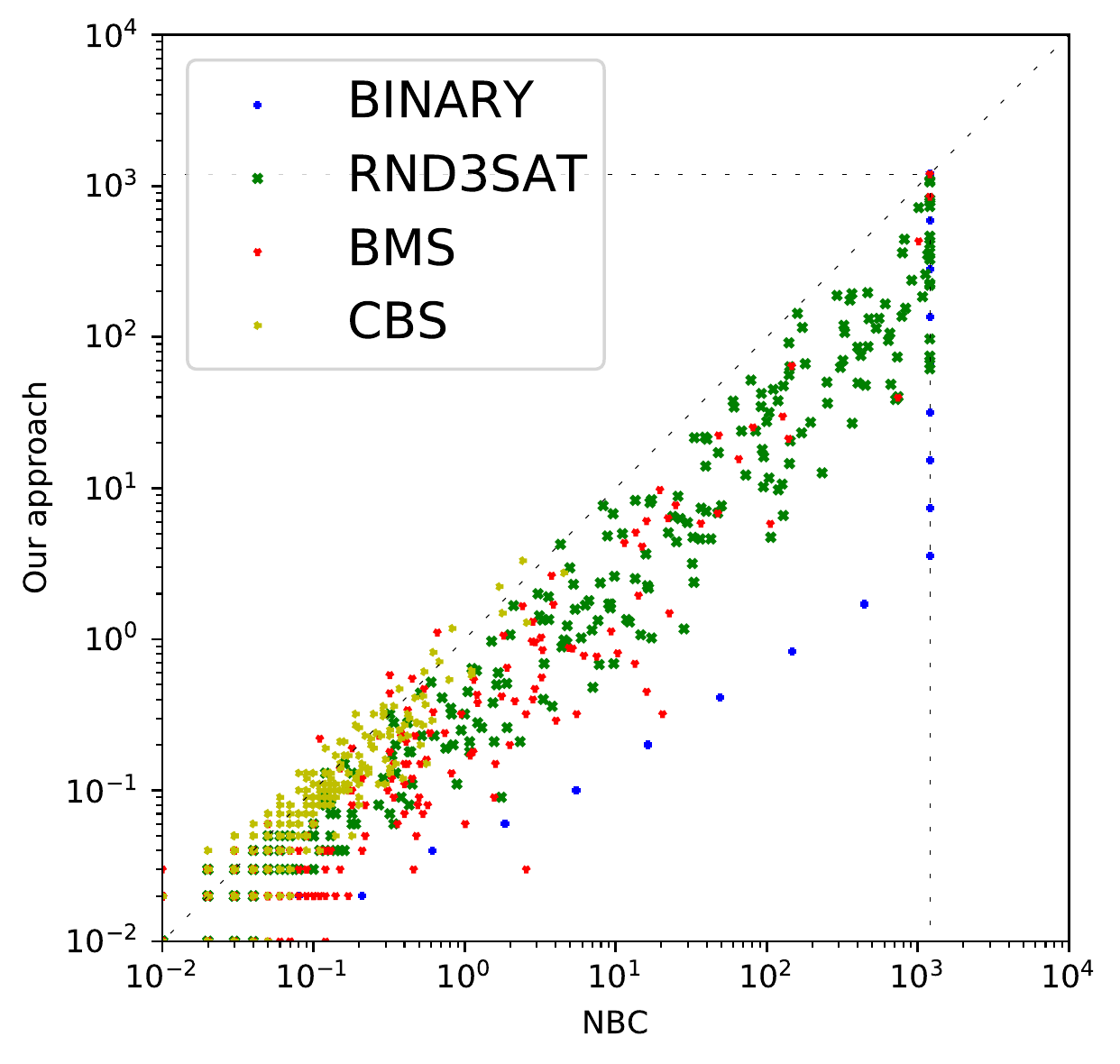}
        \caption[]%
        {{NBC}}    
        \label{fig:soa-rnd3sat}
    \end{subfigure}
    \begin{subfigure}[t]{0.19\textwidth}   
        \centering 
        \includegraphics[width=\textwidth]{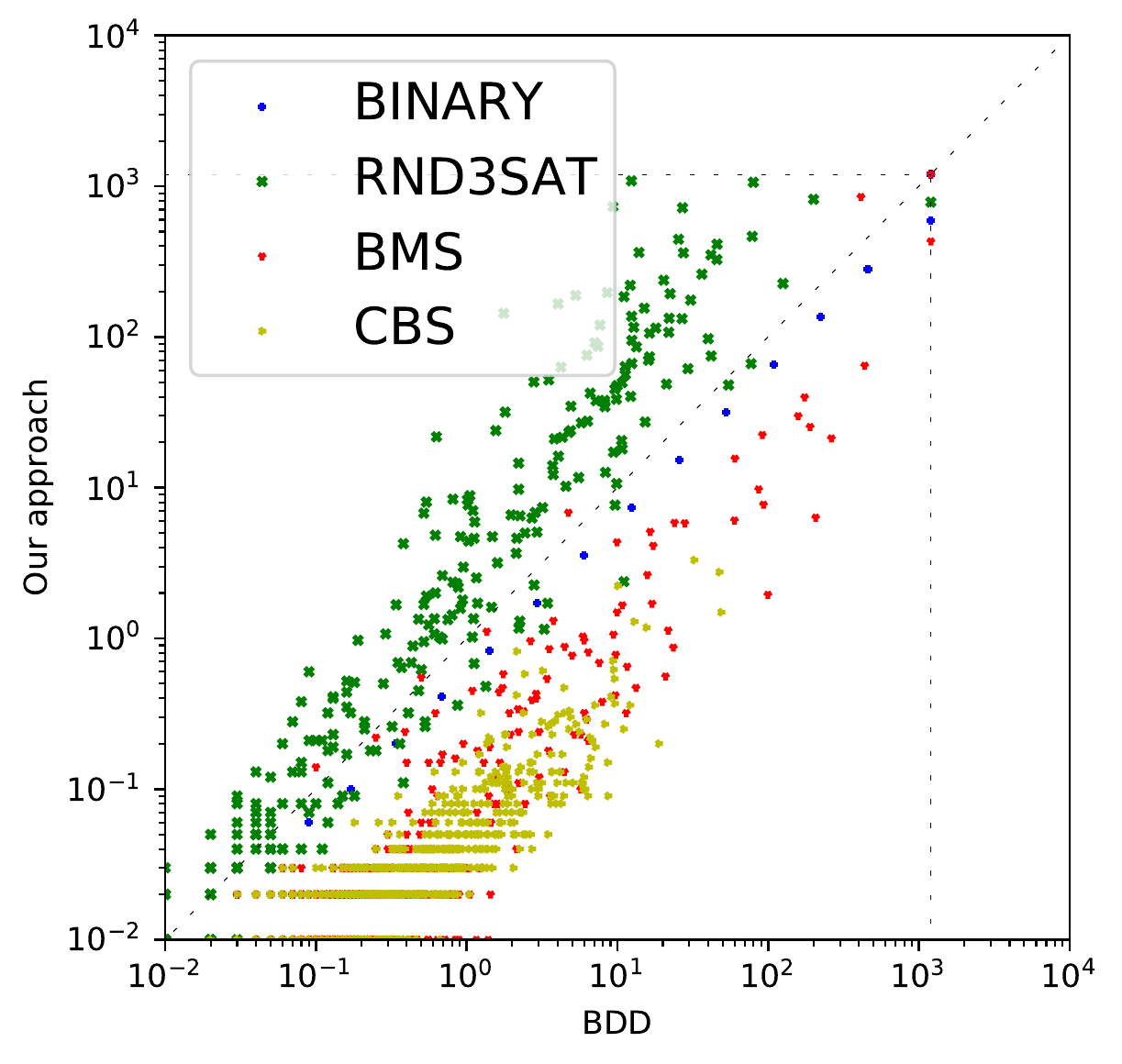}
        \caption[]%
        {{BDD}}    
        \label{fig:soa-BMSk3}
    \end{subfigure}
    \begin{subfigure}[t]{0.19\textwidth}   
        \centering 
        \includegraphics[width=\textwidth]{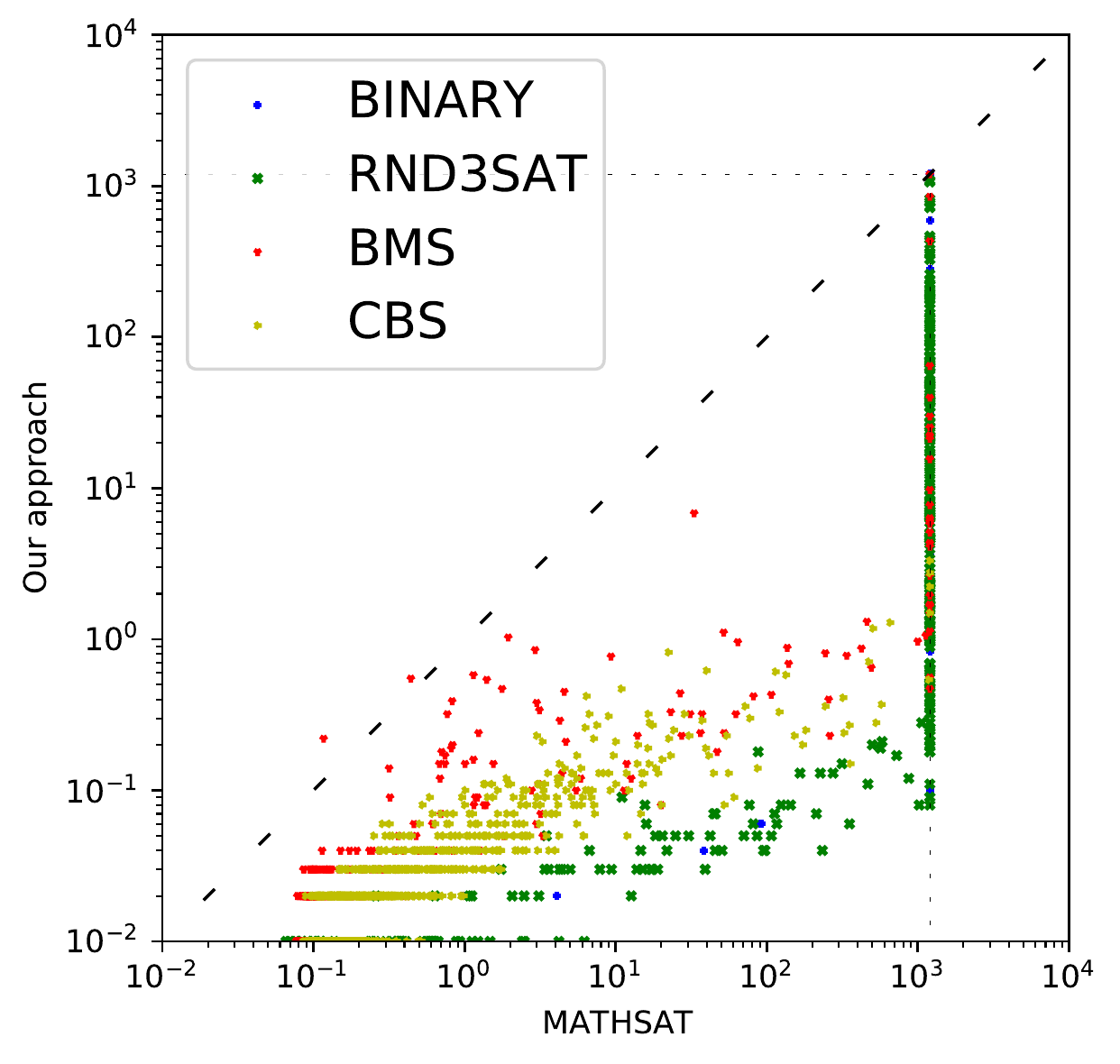}
        \caption[]%
        {{MathSAT}}    
        \label{fig:soa-newplot}
    \end{subfigure}
    \caption{Scatter plots comparing \solver{} CPU times against the other AllSAT solvers. The $x$ and $y$ axes are both log-scaled.} 
    \label{fig:stateofart}
\end{figure*}

}

\begin{figure*}[!t]
    \centering
    \begin{subfigure}[t]{0.34\textwidth}
        \centering
        \includegraphics[width=\textwidth]{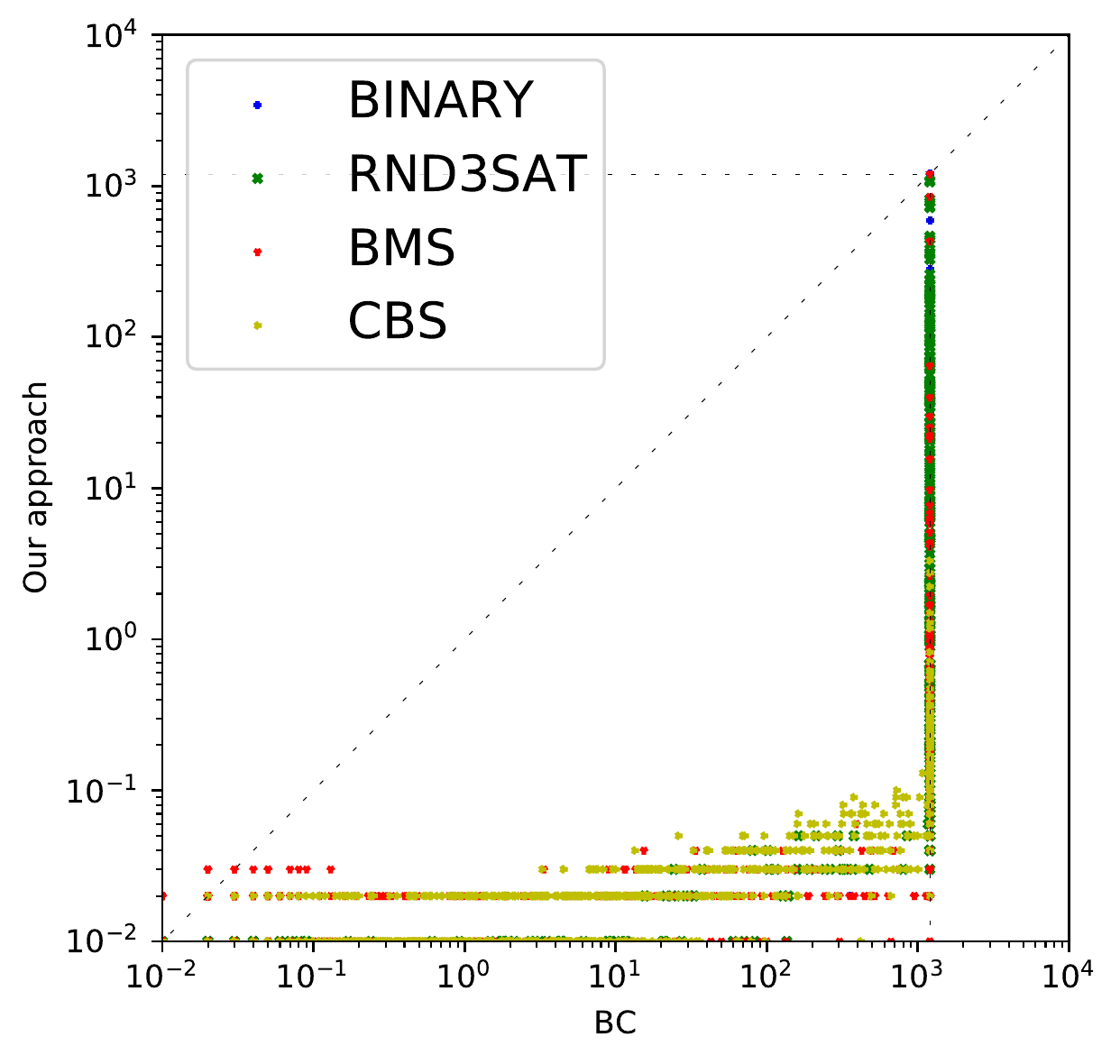}
        \caption%
        {{BC}}    
        \label{fig:soa-binary}
    \end{subfigure}
    \begin{subfigure}[t]{0.34\textwidth}  
        \centering 
        \includegraphics[width=\textwidth]{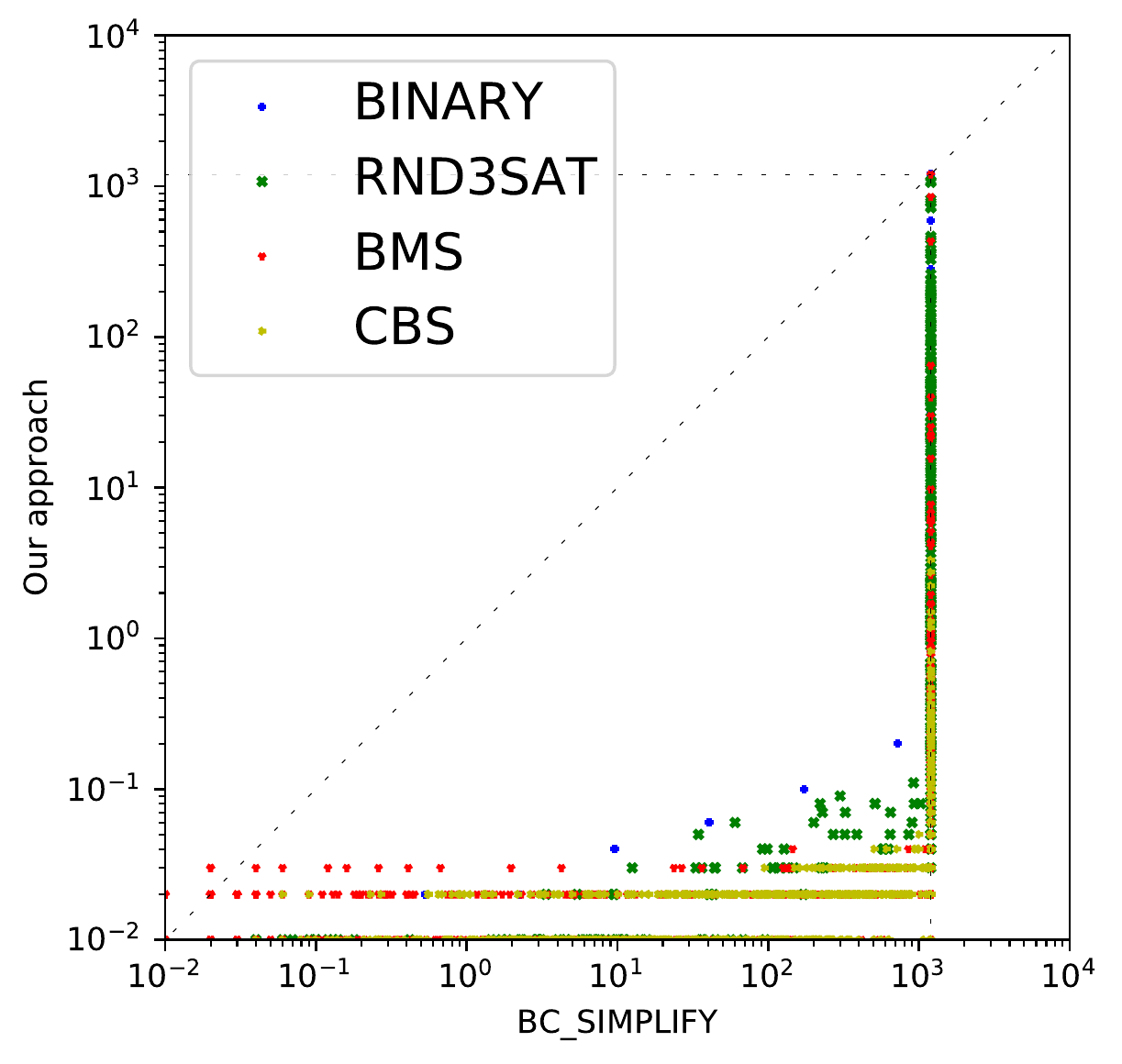}
        \caption[]%
        {{{\sc BC\_Partial}}}    
        \label{fig:soa-CSB}
    \end{subfigure}
    \begin{subfigure}[t]{0.34\textwidth}   
        \centering 
        \includegraphics[width=\textwidth]{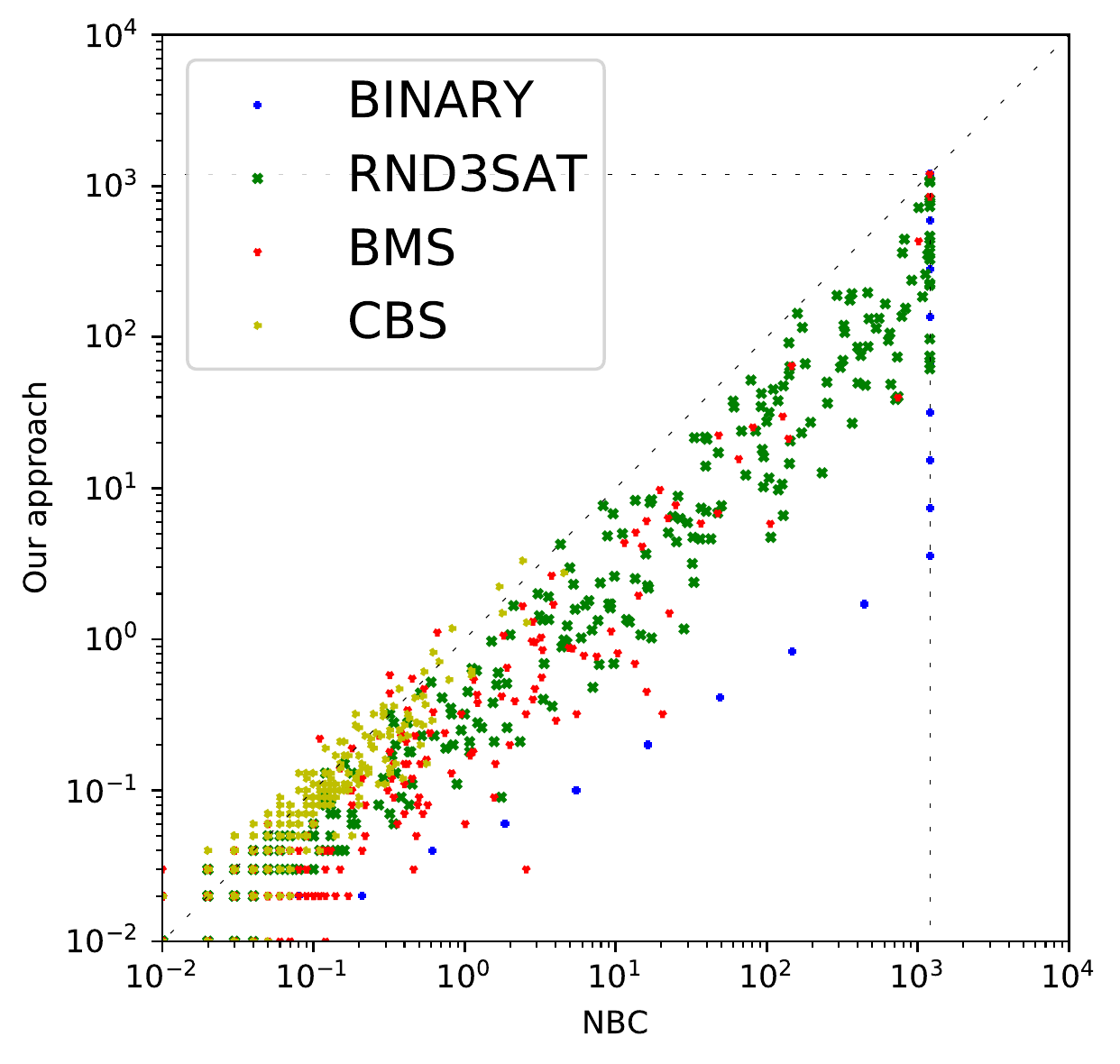}
        \caption[]%
        {{NBC}}    
        \label{fig:soa-rnd3sat}
    \end{subfigure}
    \begin{subfigure}[t]{0.34\textwidth}   
        \centering 
        \includegraphics[width=\textwidth]{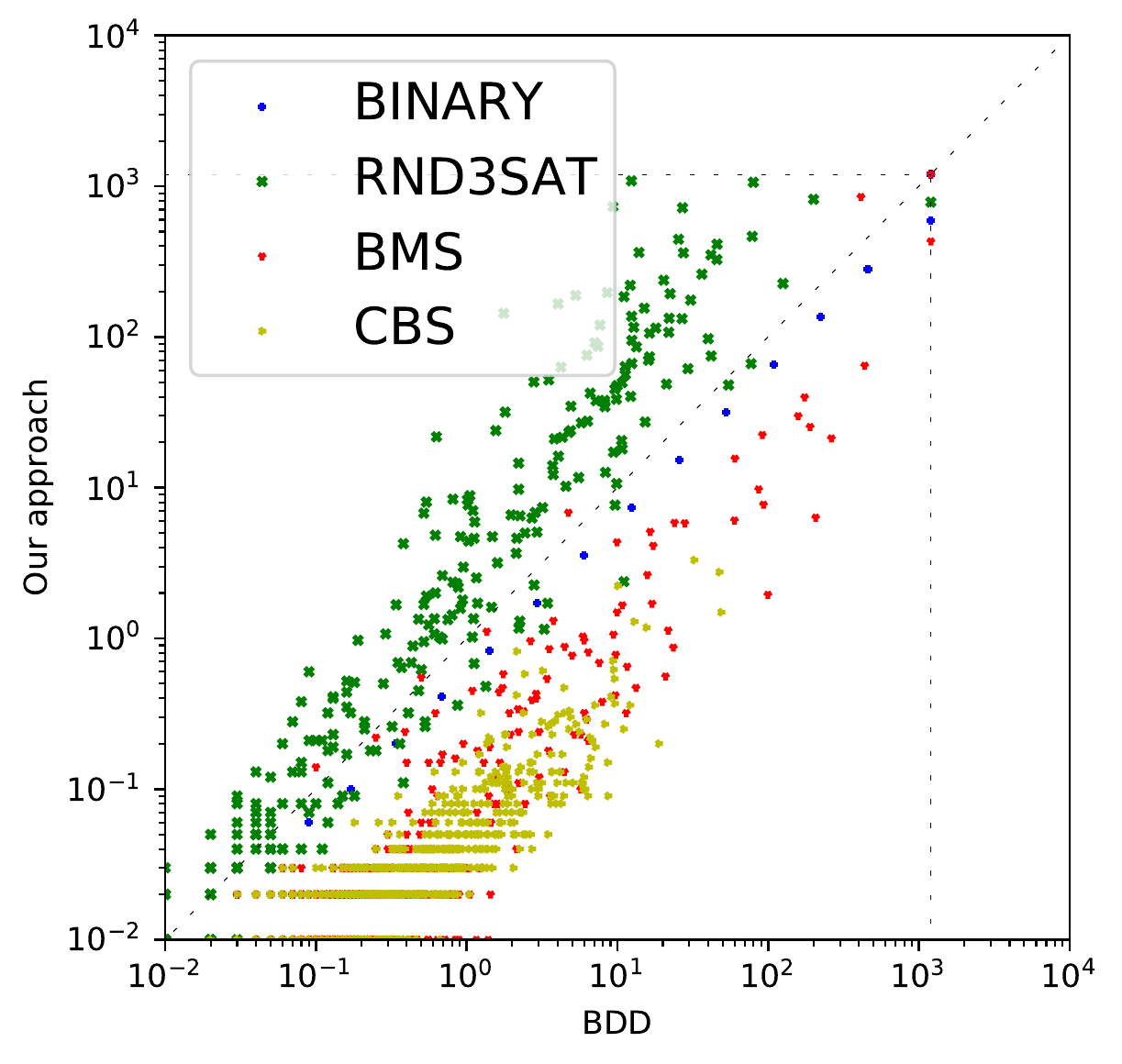}
        \caption[]%
        {{BDD}}    
        \label{fig:soa-BMSk3}
    \end{subfigure}
    \begin{subfigure}[t]{0.34\textwidth}   
        \centering 
        \includegraphics[width=\textwidth]{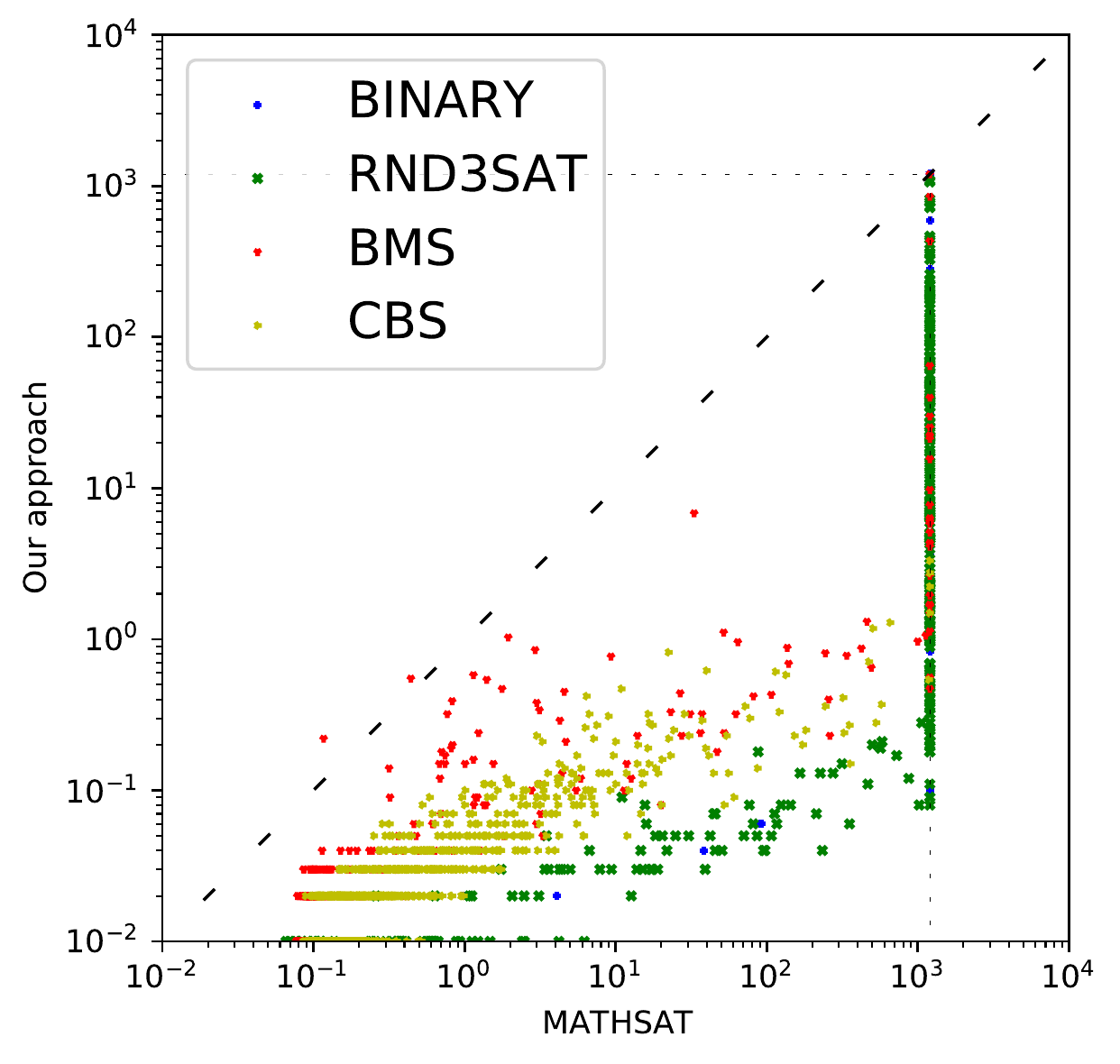}
        \caption[]%
        {{MathSAT}}    
        \label{fig:soa-newplot}
    \end{subfigure}
    \caption{Scatter plots comparing \solver{} CPU times against the other AllSAT solvers. The $x$ and $y$ axes are both log-scaled.} 
    \label{fig:stateofart}
\end{figure*}

\GSCHANGE{The benchmarks used on related works on enumeration \cite{toda2016implementing} are typically from SATLIB \cite{hoos2000satlib}, which were thought for SAT solving. However, most of these benchmarks are not suited for AllSAT solving: some benchmarks are UNSAT or admit only a couple of solutions, whereas others are encoded in a way that no total assignment can be shrunk into a partial one.}
For the sake of significance for AllSAT, we considered benchmarks having two characteristics: ($i$) each problem admits a high number of total assignments; ($ii$) the problem structure allows for some minimization of assignments, to test the efficiency of the chronological implicant shrinking algorithms.

\textit{Binary clauses} is a crafted dataset containing problems with $n$ variables defined by binary clauses in the form:
\begin{equation*}
    (x_1 \vee x_n) \wedge (x_2 \vee x_{n-1}) \wedge ... \wedge (x_{n/2-1} \vee x_{n/2})
\end{equation*}

Finding all solutions poses a significant challenge: retrieving all possible assignments requires returning $3^{n/2}$ assignments within a feasible timeframe.

\textit{Rnd3sat} contains 410 random 3-SAT problems with $n$ variables, $n\in[10,50]$. In SAT instances, the ratio of clauses to variables needed to achieve maximum hardness is about 4.26, but in AllSAT, it should be set to approximately 1.5 \cite{bayardo1997using}. For this reason, we choose not to use the instances uploaded to SATLIB and we created new random 3-SAT problems accordingly.

We also tested our algorithms over SATLIB benchmarks, specifically \textit{CBS} and \textit{BMS} \cite{singer2000backbone}. 

\subsection{Comparing Implicant Shrinking Techniques}

In Figure \ref{fig:scatter-time} we compare the two implicant shrinking algorithms with respect to CPU time and the number of disjoint partial assignments. We checked the correctness of the enumeration by testing if the number of total assignments covered by the set of partial solutions was the same as the model count reported by the \#SAT solver Ganak \cite{SRSM19}, being always correct for both algorithms.

\GSCHANGE{Results suggest that, with no surprise, dynamically updating watches is more effective in shrinking total assignments. When considering time efficiency, however, the faster but conservative simplification algorithm outperforms the other variant. The computational cost of updating each watch list $\omega(\ell)$ significantly slows down the computation process the higher the number of total models satisfying $F$ is.}

All the experiments in the following subsections assume \solver{} relies on the lighter variant.

\subsection{Baseline Solvers}

We considered BC, NBC, and BDD \cite{toda2016implementing}, respectively a blocking, a non-blocking, and a BDD-based disjoint AllSAT solver. BC also provides the option to obtain partial assignments (from now on \textsc{BC\_Partial}). Lastly, we considered \textsc{MathSAT5} \cite{cimatti2013mathsat5}, since it provides an interface to compute partial enumeration of propositional problems by exploiting blocking clauses.

\GMCHANGE{
Some other AllSAT solvers, such as {\sc BASolver} \cite{zhang2020accelerating} and {\sc AllSATCC} \cite{liang2022allsatcc}, are currently not publicly available, as reported also in another paper \cite{fried2023allsat}.
}

\subsection{Results}

Table \ref{tb:table} reports the number of instances solved by each solver for each set of benchmarks before reaching timeout, \GSCHANGE{where "solved" means that they enumerated completely a set of disjoint partial models covering all total models}. \GMCHANGE{We see that \solver{} solves the highest amount of instances for each benchmark, even though {\sf BDD} and {\sf NBC} are close. We also present some scatter plots comparing \solver{} time performance against each of the other AllSAT solvers available, using different marks and colors to distinguish instances from different benchmarks. The CPU times reported in Figure \ref{fig:stateofart} consider only the time taken to reach each assignment, without storing them.} 

\solver{} outperforms all the other %baseline 
solvers in %almost %RS: se dici except, non occorre dire almost
every benchmark excluding \textsc{rnd3sat}, where BDD 
%slightly % RS: not slightly
outperforms our approach. The latter instances are not structurally complex due to the low clause-to-variable ratio and can be compiled into BDDs with minimal inefficiencies, thus justifying this behavior: the higher the number of clauses is,
%in the problem instance, 
the more challenging the compilation of the propositional formula into a BDD is,
as we can see with BMS and CSB.

\section{Conclusion}
\label{sec:conclusion}

We presented an AllSAT procedure that combines CDCL, CB, and chronological implicant shrinking to perform partial disjoint enumeration. The experiments confirm the benefits of 
combining them, avoiding both performance degradations due to blocking clauses and bottlenecks generated by the solver being stuck in non-satisfiable search sub-trees.

\GSCHANGE{This work could be extended in several directions. First, we plan to compare our algorithm against other enumeration algorithms based on knowledge compilation (for instance {\sc D4} \cite{lagniez2017improved}), even though this might involve a potentially costly compilation process before enumeration and accordingly such an approach is not any-time. Then, to further improve the performances of \solver{},
we plan to explore novel decision heuristics
that are suitable for chronological backtracking. Finally, we plan
to extend our techniques to handle also {\em projected enumeration}
and to investigate the integration of chronological backtracking with
component caching.}

\section*{Acknowledgements}

\GSCHANGEBIS{
We acknowledge the support of the MUR PNRR project FAIR --
  Future AI Research (PE00000013), under the NRRP MUR program funded
  by the NextGenerationEU. The work was partially supported by the
  project ``AI@TN'' funded by the Autonomous Province of Trento. This research was partially supported by TAILOR, a project funded
by the EU Horizon 2020 research and innovation program under GA No 952215.
}

\bibliography{aaai24}

\end{document}